\documentclass[a4paper,11pt]{article}
\usepackage{jinstpub}

\usepackage{graphicx}
\usepackage{subfig}
\usepackage{siunitx}
\usepackage{booktabs}
\usepackage{titlesec}
\usepackage{lineno}

\titlespacing*{\section}
  {0pt}{1.2ex}{0.6ex}

\title{\boldmath Investigation of the timing performance
of double-sided $p$-type 3D silicon sensors produced by CNM}

\author[a,1]{C.~Liao\note{Corresponding author.}}
\author[b]{C.~Betancourt}
\author[a,c]{M.~Togawa}
\author[a,d]{D.~Bortoletto}
\author[e]{N.~Moffat}
\author[e]{and G.~Pellegrini}

\affiliation[a]{
International Center for Quantum-field Measurement Systems
for Studies of the Universe and Particles (QUP, WPI),
High Energy Accelerator Research Organization (KEK),\\
Oho 1-1, Tsukuba, Ibaraki 305-0801, Japan}

\affiliation[b]{
SINTEF, Smart Sensors and Microsystems,\\
Oslo, Norway}

\affiliation[c]{
IPNS, High Energy Accelerator Research Organization (KEK),\\
Tsukuba, Japan}

\affiliation[d]{
University of Oxford,\\
Oxford, United Kingdom}

\affiliation[e]{
Centro Nacional de Microelectr\'onica (CNM), CSIC,\\
Barcelona, Spain}

\emailAdd{liao@post.kek.jp}

\abstract{
We investigate the timing performance and spatial uniformity
of double-sided $p$-type 3D silicon sensors produced by the
Centro Nacional de Microelectr\'onica (CNM).
Three sensor layouts are studied: hexagonal and square
geometries with an inter-electrode spacing of
\SI{30}{\micro\meter}, and a hexagonal geometry with
an inter-electrode spacing of \SI{50}{\micro\meter}.
The sensors are characterized using a
\SI{1064}{\nano\meter} infrared laser and a custom
two-stage wideband readout.
Measurements with a $^{90}$Sr source are used to establish
a minimum-ionizing-particle-like reference for the laser
signal.
The local timing resolution improves with increasing
bias voltage for all three layouts.
Two-dimensional laser scans are used to investigate
the spatial variations of both the local timing
resolution and the mean threshold-crossing time.
Spatial variations in the latter introduce an additional timing uncertainty
when the particle impact position is unknown.
Among the investigated devices, the square
\SI{30}{\micro\meter} geometry shows the best overall
timing performance at the highest measured bias voltage.
}


\begin{document}

\maketitle


\flushbottom

\section{Introduction}
\label{sec:introduction}

The High-Luminosity Large Hadron Collider (HL-LHC) will
operate with a nominal bunch-crossing frequency of 40~MHz and up to approximately 200 proton--proton interactions per bunch crossing. The resulting high-pileup and high-radiation environment presents significant challenges for particle tracking and event reconstruction. Although precise timing is not a baseline requirement for the inner tracking detectors, additional timing information could improve pileup mitigation and event reconstruction.
\par
Low Gain Avalanche Detectors (LGADs)~\cite{b1,b2,b3} can achieve timing resolutions below 50~ps through internal charge multiplication. However, radiation-induced acceptor removal in the gain layer can degrade their performance, despite mitigation techniques such as carbon co-implantation~\cite{b4}. In contrast, conventional
3D silicon sensors operate without an avalanche gain layer
and offer a different combination of spatial resolution,
radiation hardness, and timing capabilities.
\par
In 3D silicon sensors, the electrode spacing is decoupled
from the sensor thickness~\cite{b5}, enabling short carrier
drift distances and fast charge collection even after
high radiation fluences. These sensors have demonstrated
operation at fluences beyond $\Phi_{\mathrm{eq}}=\num{1e16}~\si{cm^{-2}}$ as well as promising timing performance~\cite{b6}. However, fast charge collection alone does not guarantee excellent timing resolution, which also depends on the induced-current pulse shape, signal-to-noise ratio, carrier transport, and the spatial distributions of the electric and weighting fields. Consequently, electrode geometry and spacing may
influence both the local timing resolution and
the mean signal response time.
Spatial variations in the latter introduce an
additional timing uncertainty when the particle
impact position is unknown.

In this work, we investigate the timing performance of
double-sided $p$-type 3D silicon sensors using a
$^{90}$Sr source and a \SI{1064}{\nano\meter}
infrared laser.
Three layouts are compared to evaluate the dependence
of the local timing resolution on bias voltage and
the influence of electrode spacing and geometry.
Furthermore, the spatial variations of the local
timing resolution and the mean threshold-crossing
time are investigated to assess the additional
timing uncertainty associated with unknown particle
impact positions.
The results provide experimental guidance for
optimizing future 3D sensors with timing capabilities.

\section{Sensors}
\label{sec:sensors}



\begin{figure}[htbp]
  \centering

  \subfloat[\label{fig:schematic}]{%
    \includegraphics[width=0.90\linewidth]
    {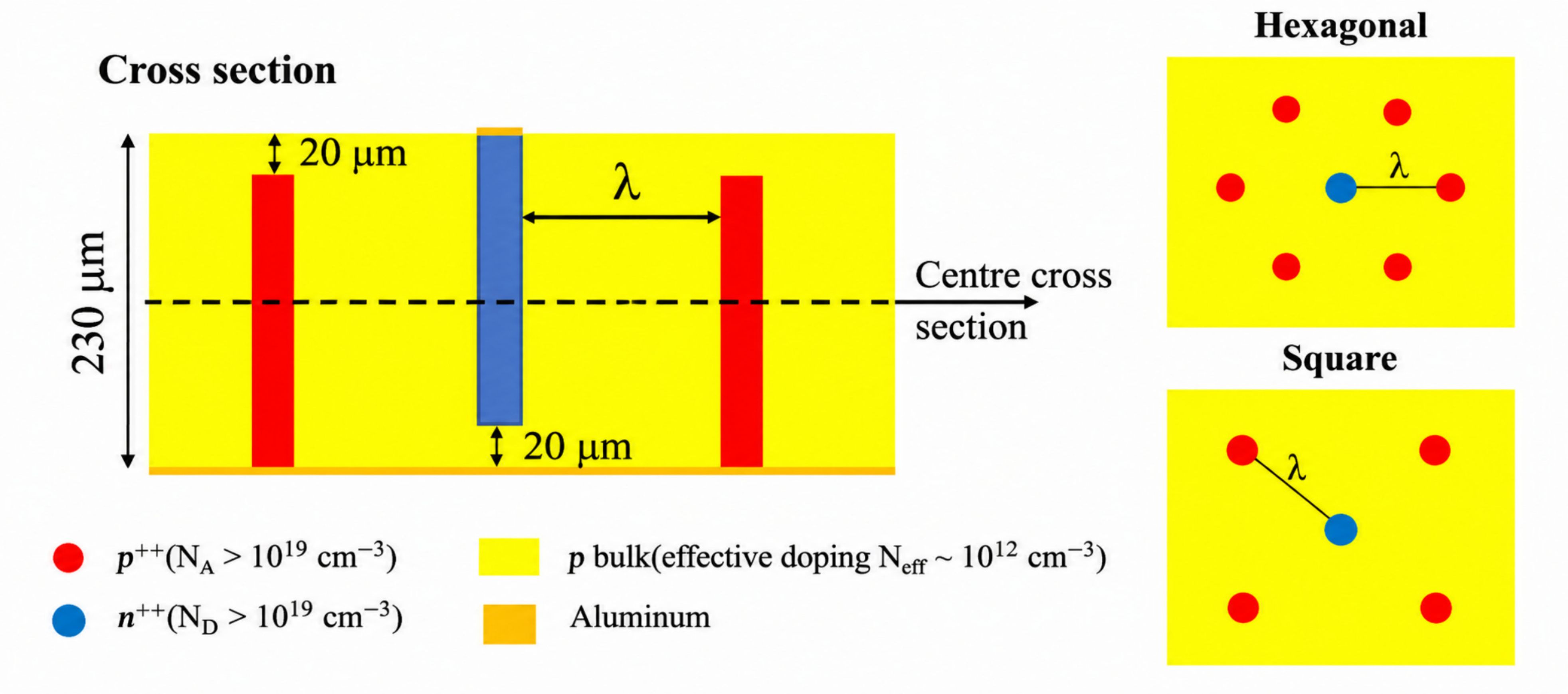}
  }

  \par\medskip

  \subfloat[\label{fig:sensor_hex30}]{%
    \includegraphics[width=0.3\linewidth]
    {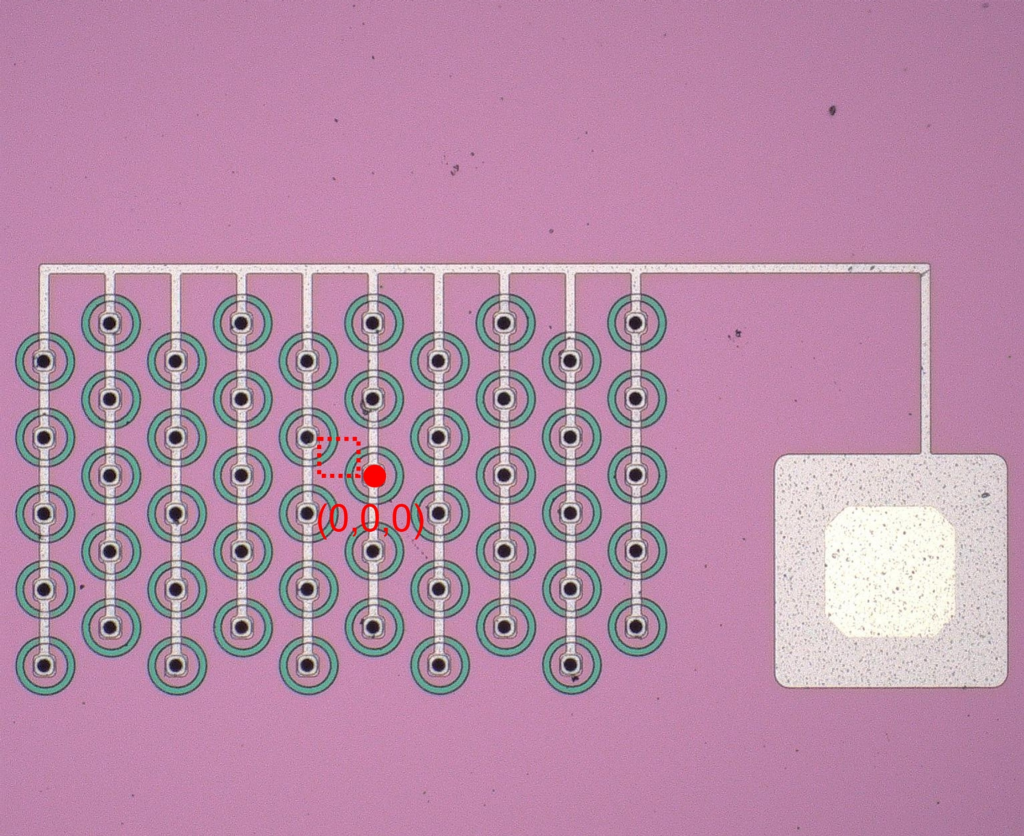}
  }
  \hfill
  \subfloat[\label{fig:sensor_square30}]{%
    \includegraphics[width=0.3\linewidth]
    {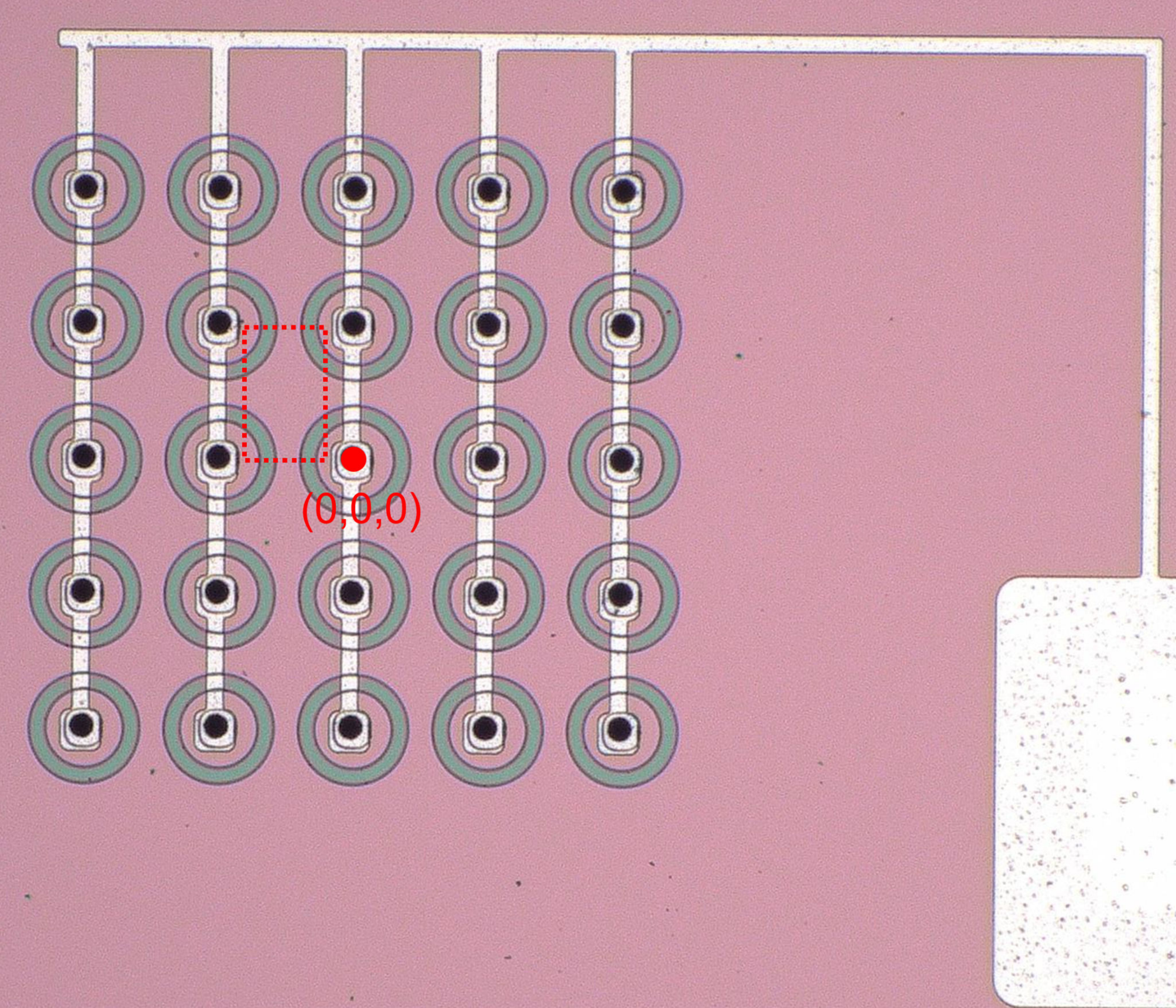}
  }
  \hfill
  \subfloat[\label{fig:sensor_hex50}]{%
    \includegraphics[width=0.3\linewidth]
    {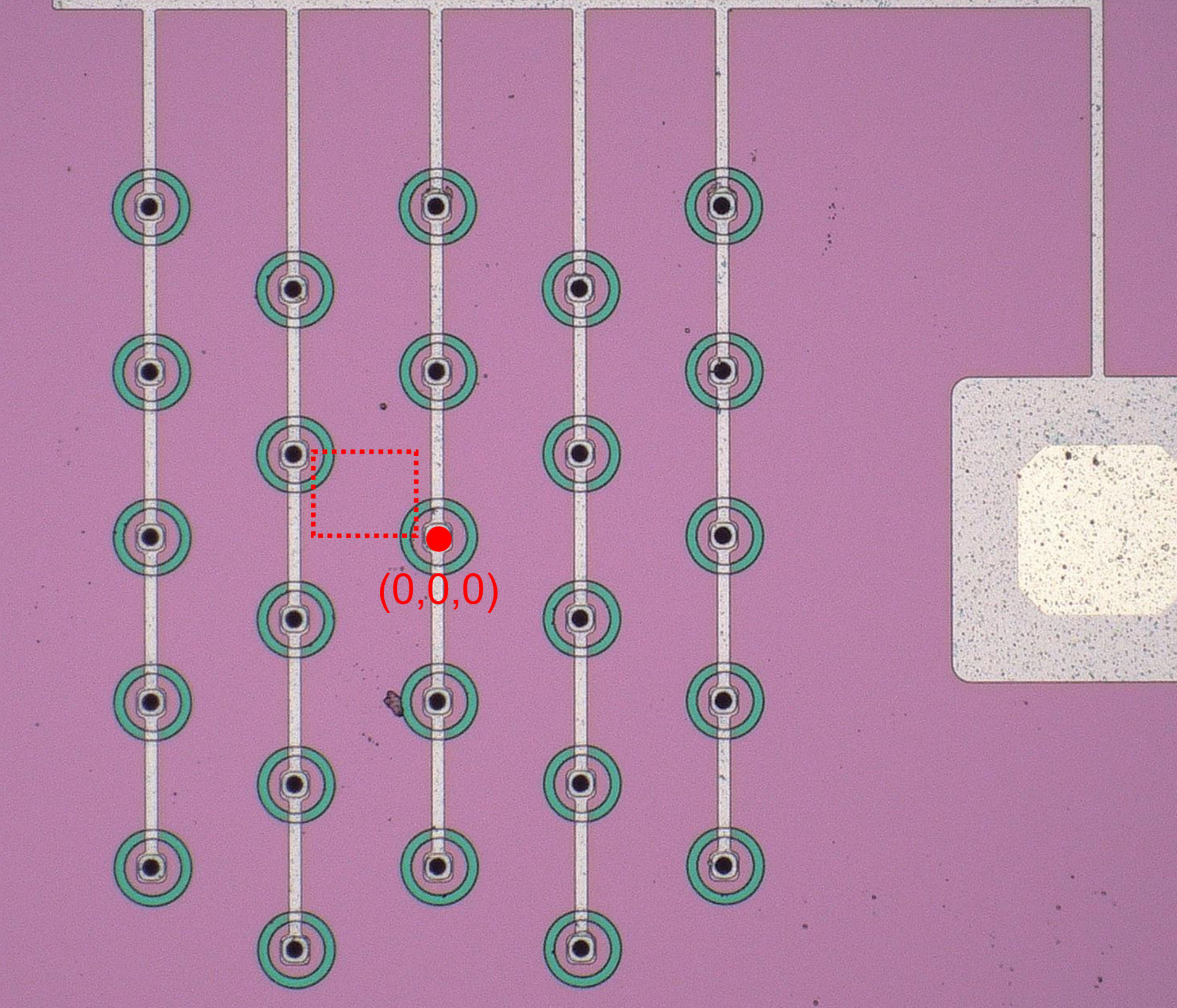}
  }

  \caption{Schematic illustration and front-side optical
  images of the investigated double-sided 3D silicon
  sensors.
  (a) Cross-sectional view showing the $n^{+}$ and
  $p^{+}$ electrode configurations.
  (b) Hexagonal geometry with a nominal inter-electrode
  spacing of \SI{30}{\micro\meter} (Hex30).
  (c) Square geometry with a nominal inter-electrode
  spacing of \SI{30}{\micro\meter} (Square30).
  (d) Hexagonal geometry with a nominal inter-electrode
  spacing of \SI{50}{\micro\meter} (Hex50).
  The red boxes indicate the regions selected for
  the infrared laser scans.}

  \label{fig:sensors}
\end{figure}


\begin{table}[htbp]
  \centering
  \caption{Main parameters of the investigated
  double-sided $p$-type 3D silicon sensors.}
  \label{tab:sensor_parameters}

  \small
  \setlength{\tabcolsep}{5pt}
  \renewcommand{\arraystretch}{1.15}

  \begin{tabular}{lccc}
    \toprule
    Parameter & Hex30 & Square30 & Hex50 \\
    \midrule
    Electrode geometry
      & Hexagonal & Square & Hexagonal \\
    Inter-electrode spacing ($\mu$m)
      & 30 & 30 & 50 \\
    Column depth ($\mu$m)
      & 210 & 210 & 210 \\
    Cell matrix
      & $5\times10$ & $5\times5$ & $5\times5$ \\
    Column diameter ($\mu$m)
      & 10 & 10 & 10 \\
    \bottomrule
  \end{tabular}
\end{table}


The investigated sensors were fabricated by the
Centro Nacional de Microelectr\'onica (CNM)~\cite{b7}
using high-resistivity Czochralski (Cz) silicon
wafers with a thickness of \SI{230}{\micro\meter}.
The devices employ a double-sided 3D configuration,
in which the $n^{+}$ and $p^{+}$ electrode columns
are etched from opposite sides of the substrate
without extending through its entire thickness,
as illustrated in Figure~\ref{fig:schematic}.
Further fabrication details can be found in
Ref.~\cite{b7}.

\par

Three electrode layouts (Hex30, Square30, and Hex50)
were investigated, as shown in Figure~\ref{fig:sensors}.
The readout electrodes are connected to a common
pad through aluminum metallization.
The main device parameters are summarized in
Table~\ref{tab:sensor_parameters}.

\par

The nominal inter-electrode spacing is defined as
the center-to-center distance between neighboring
$n^{+}$ and $p^{+}$ electrode columns. It was
estimated from optical microscopy images based
on the periodic electrode geometry.
The backside $p^{+}$ electrode columns are not
visible in the front-side optical images.

\section{Timing setup}
\label{sec:timing_setup}

\begin{figure}[htbp]
  \centering

  \subfloat[Laser measurement setup.
  \label{fig3a}]{%
    \includegraphics[width=0.6\linewidth]
    {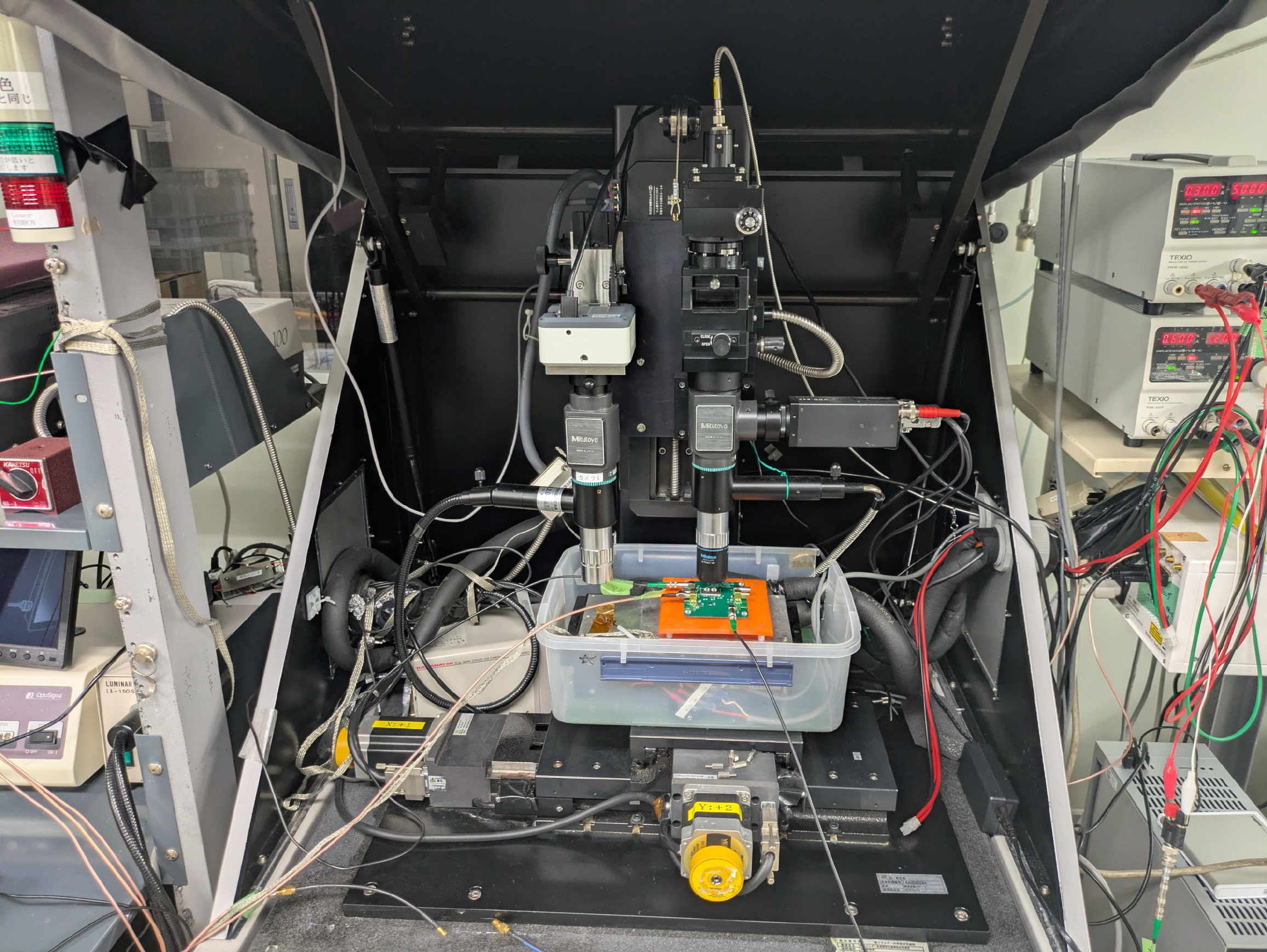}}
  \hfill
  \subfloat[Optical monitoring system.
  \label{fig3b}]{%
    \includegraphics[width=0.3\linewidth]
    {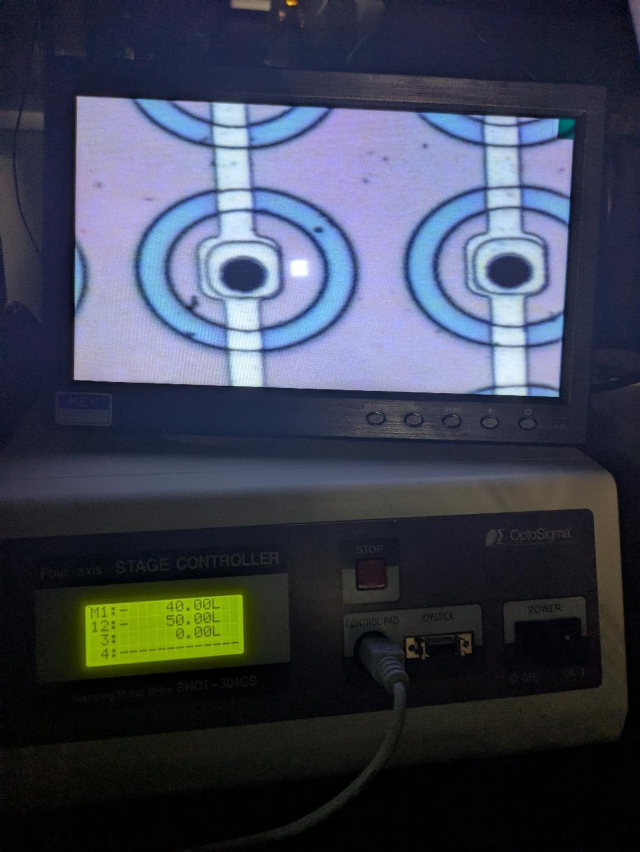}}

  \caption{Laser measurement setup and optical
  monitoring system used for laser alignment
  and spatial timing measurements.}

  \label{fig3}
\end{figure}

Timing measurements were performed at room temperature
using a picosecond-pulsed infrared laser (NKT Photonics)
with a wavelength of \SI{1064}{\nano\meter}, a repetition
rate of \SI{1}{\mega\hertz}, and a focused spot size of
approximately \SI{2}{\micro\meter}$\times$
\SI{2}{\micro\meter}. The laser intensity was calibrated
against $^{90}$Sr signals to establish a MIP-like reference.
The experimental setup is shown in Figure~\ref{fig3}.
The central $n^{+}$ column was defined as the origin
$(0,0,0)$ for the spatial scans.

\par

The detector signals were recorded using a LeCroy
WavePro 804HD oscilloscope with an analog bandwidth
of \SI{8}{\giga\hertz}. The oscilloscope was
triggered by the laser synchronization output.
The recorded waveforms were used to extract the
timing resolution and the mean threshold-crossing
time $\mu$, as described in
Section~\ref{sec:timing_analysis}.
The timing-reference jitter of the laser system
is therefore included in the measured timing resolution.

\par

For a fixed laser injection position, the
constant-fraction method suppresses amplitude-dependent time walk. Assuming negligible correlations between the individual contributions, the measured local timing resolution can be approximated as~\cite{b11}

\begin{equation}
\sigma_{\mathrm{local}}^{2}
\simeq
\sigma_{\mathrm{sensor}}^{2}
+
\sigma_{\mathrm{jitter}}^{2}
+
\sigma_{\mathrm{ref}}^{2},
\label{eq:timing_resolution}
\end{equation}

where $\sigma_{\mathrm{sensor}}$ represents intrinsic
event-to-event fluctuations in the sensor response,
including fluctuations in the signal pulse shape,
$\sigma_{\mathrm{jitter}}$ is the noise-induced
timing jitter, and $\sigma_{\mathrm{ref}}$ is the
timing-reference jitter.

The electronic timing jitter can be approximated as

\begin{equation}
\sigma_{\mathrm{jitter}}
\simeq
\frac{\sigma_V}
{\left|\mathrm{d}V/\mathrm{d}t\right|_{\mathrm{th}}},
\label{eq:electronic_jitter}
\end{equation}

where $\sigma_V$ is the RMS output voltage noise and
$\left|\mathrm{d}V/\mathrm{d}t\right|_{\mathrm{th}}$
is the absolute signal slope at the threshold-crossing
point.

\par

Position-dependent variations in carrier transport
and induced-current pulse shapes can lead to
systematic shifts in the mean threshold-crossing
time $\mu$.
These shifts do not contribute to the local timing
resolution at a fixed injection position but
introduce an additional timing uncertainty when
the particle impact position is unknown and no
position-dependent timing correction is applied.
This contribution is investigated in
Section~\ref{sec:spatial_uniformity}.

\par

A timing-reference jitter of approximately
\SI{5}{\pico\second} was reported for the
infrared laser setup used in previous
measurements~\cite{nakamura_laser}.
This contribution was not independently measured
in the present study and is included in the
measured local timing resolution.
The output voltage noise relevant to
$\sigma_{\mathrm{jitter}}$ is characterized
in the following section.

\section{Readout system}
\label{sec:readout}

\begin{figure}[htbp]
    \centering

    \subfloat[\label{fig:readout_schematic}]{%
        \includegraphics[width=0.85\linewidth]
        {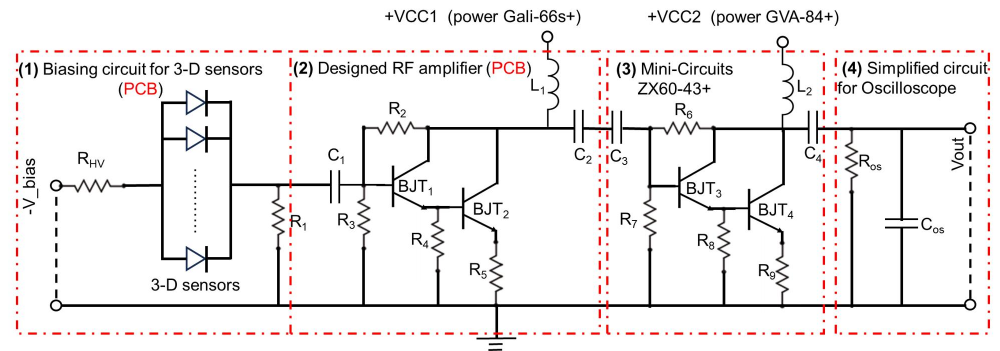}
    }

    \par\medskip

    \subfloat[\label{fig:amplification}]{%
        \includegraphics[width=0.48\linewidth]
        {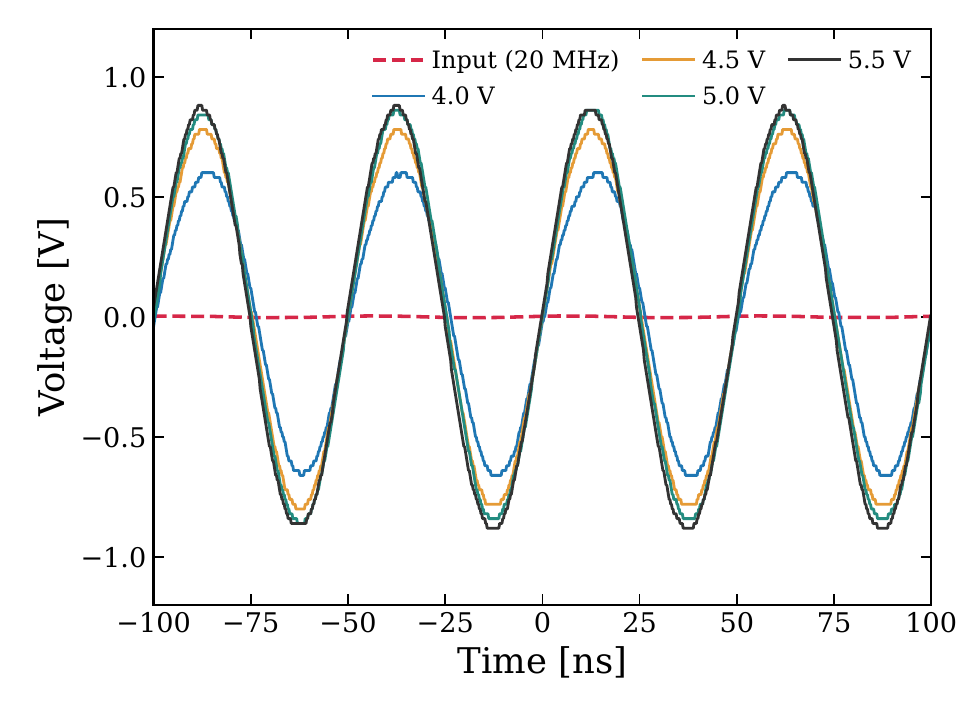}
    }
    \hfill
    \subfloat[\label{fig:readout_histogram}]{%
        \includegraphics[width=0.48\linewidth]
        {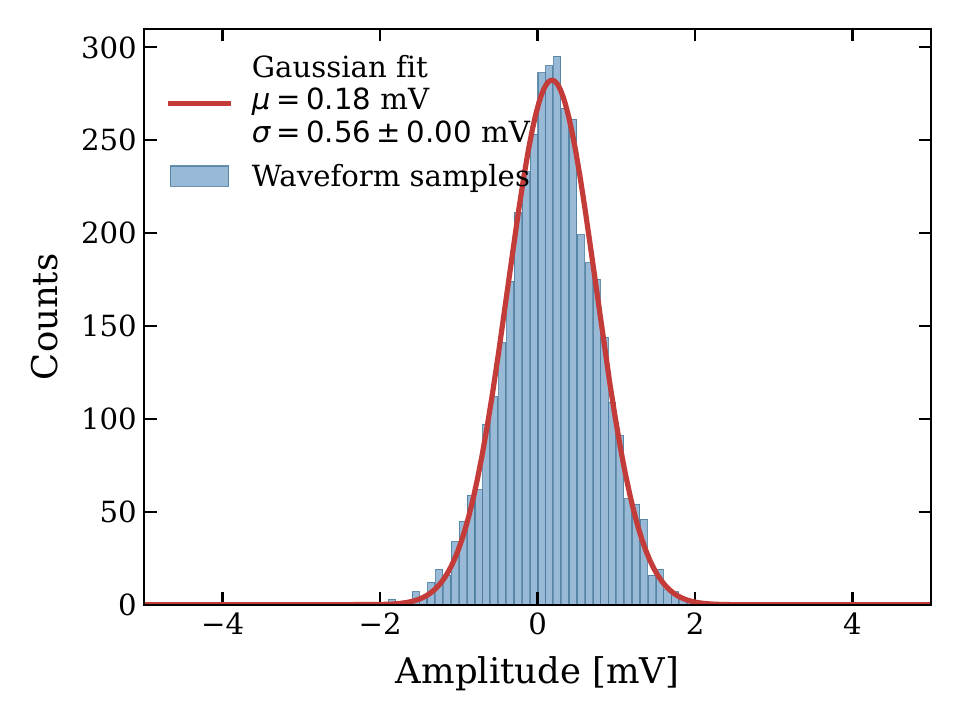}
    }

    \par\medskip

    \subfloat[\label{fig:noise_bias}]{%
        \includegraphics[width=0.48\linewidth]
        {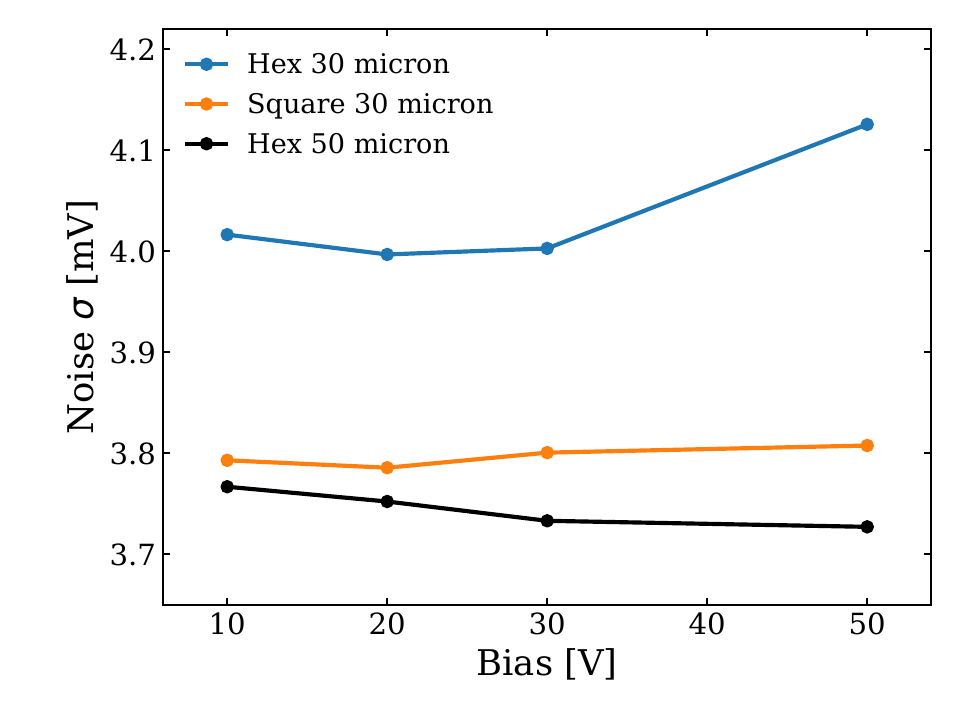}
    }
    \hfill
    \subfloat[\label{fig:iv}]{%
        \includegraphics[width=0.48\linewidth]
        {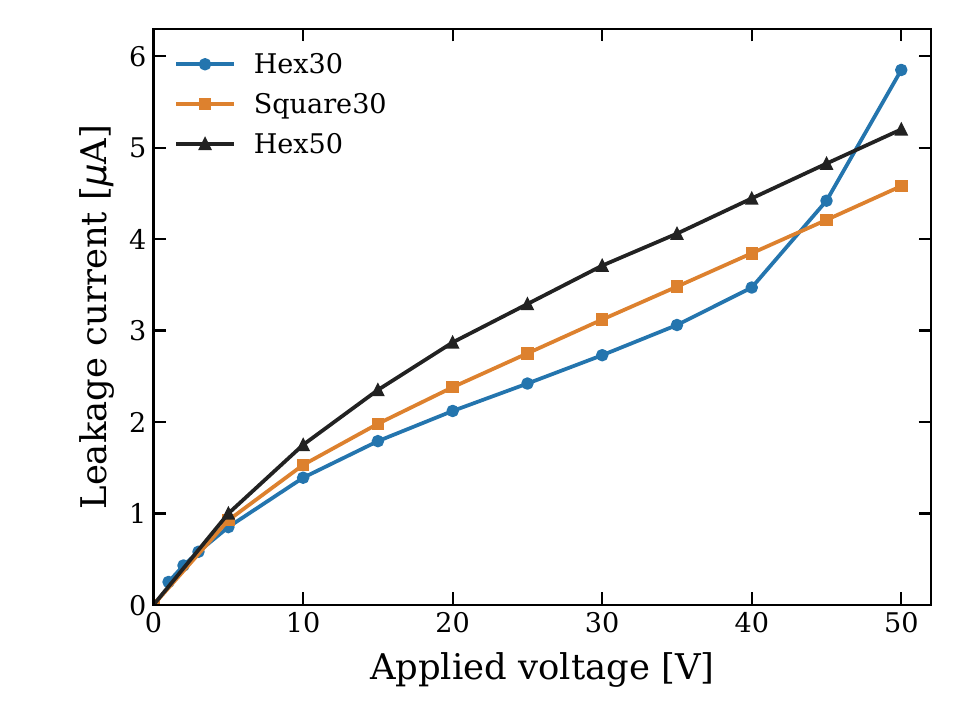}
    }

    \caption{Electrical characterization of the readout
    system and the three investigated 3D sensors.
    (a) Schematic of the two-stage wideband readout
    and detector biasing circuit.
    (b) Input and output waveforms measured at
    different second-stage supply voltages.
    (c) Gaussian fit to the voltage distribution of
    a representative waveform recorded under
    open-input conditions.
    (d) RMS output voltage noise as a function of
    bias voltage with the sensors connected.
    (e) Detector leakage current as a function
    of applied voltage.}

    \label{fig:readout_char}
\end{figure}

For the timing measurements, a custom two-stage wideband
readout system was employed. The first stage consists of
a PCB-based transimpedance preamplifier using a Gali-66S+
device, with a reported bandwidth exceeding
\SI{4}{\giga\hertz}. The second stage uses a commercial
wideband amplifier (Mini-Circuits ZX60-43+) to further
amplify the detector signal for digitization. The readout
and detector biasing scheme is illustrated in
Figure~\ref{fig:readout_char}(a). Further details of the
readout system are provided in Ref.~\cite{b8}.

\par

The overall voltage gain was characterized using a
\SI{20}{\mega\hertz} sinusoidal test signal. With the
first-stage supply fixed at \SI{5}{\volt}, the measured
gain increased from approximately 172 to 238~V/V as
the second-stage supply voltage increased from
\SI{4.0}{\volt} to \SI{5.5}{\volt}, as shown in
Figure~\ref{fig:readout_char}(b).

\par

The electronics-only RMS output voltage noise was
evaluated using more than 2000 waveforms recorded
with the detectors removed and the test-signal
generator disconnected. Both signal-input connections
before the coupling capacitor were left open, while
the two amplifier stages were powered at
\SI{5}{\volt}. For each waveform, a histogram of
the recorded voltage samples was constructed and
fitted with a Gaussian function to extract the
standard deviation $\sigma_V$, as illustrated in
Figure~\ref{fig:readout_char}(c). The resulting
$\sigma_V$ distribution was fitted with a second
Gaussian function, whose mean yielded a
representative electronics-only RMS output voltage
noise of approximately \SI{0.5}{\milli\volt}.

\par

The RMS output voltage noise with the sensors
connected and biased was evaluated using the same
procedure. More than 2000 laser-triggered waveforms
were recorded at a fixed position on each sensor.
For each waveform, the Gaussian fit was restricted
to the baseline-dominated peak of the voltage
distribution. As shown in
Figure~\ref{fig:readout_char}(d), the measured
RMS output voltage noise was approximately
\SI{4}{\milli\volt} for all three sensor geometries,
substantially higher than the electronics-only
value measured under open-input conditions.

\par

Although the measured RMS voltage noise is comparable
among the three sensors, differences in sensor
capacitance can affect the leading-edge slope and
thus the electronic timing jitter, according to
Equation~\ref{eq:electronic_jitter}.
The individual noise contributions from the
sensor and readout electronics could not be separated under
operating conditions.

\par

The detector bias was supplied through two
\SI{1}{\mega\ohm} series resistors, as illustrated
in Figure~\ref{fig:readout_char}(a). The effective
sensor bias voltage was calculated by correcting
the applied voltage for the voltage drop across
these resistors using the measured detector
leakage current:

\begin{equation}
V_{\mathrm{bias}}
=
V_{\mathrm{applied}}
-
I_{\mathrm{leak}}(R_{\mathrm{HV}}+R_1),
\label{eq:bias_correction}
\end{equation}

where $I_{\mathrm{leak}}$ is the measured detector
leakage current at the corresponding applied
voltage, and
$R_{\mathrm{HV}}=R_1=\SI{1}{\mega\ohm}$.
The measured leakage currents for the three
sensor geometries are shown in
Figure~\ref{fig:readout_char}(e) as a function
of the applied voltage.

\section{Laser calibration}
\label{sec:measurements}

\begin{figure}[htbp]
    \centering

    \subfloat[\label{fig:beta_waveforms}]{%
        \includegraphics[width=0.43\linewidth]{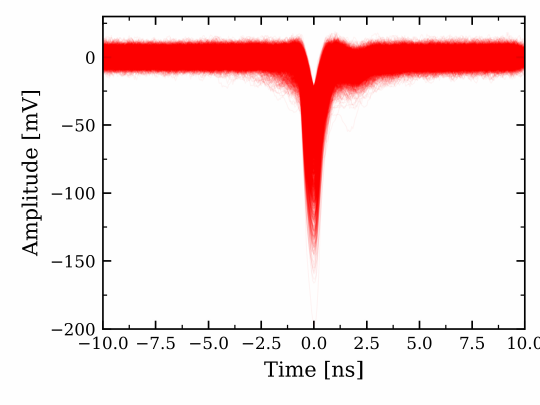}
    }
    \hfill
    \subfloat[\label{fig:beta_fit}]{%
        \includegraphics[width=0.43\linewidth]
        {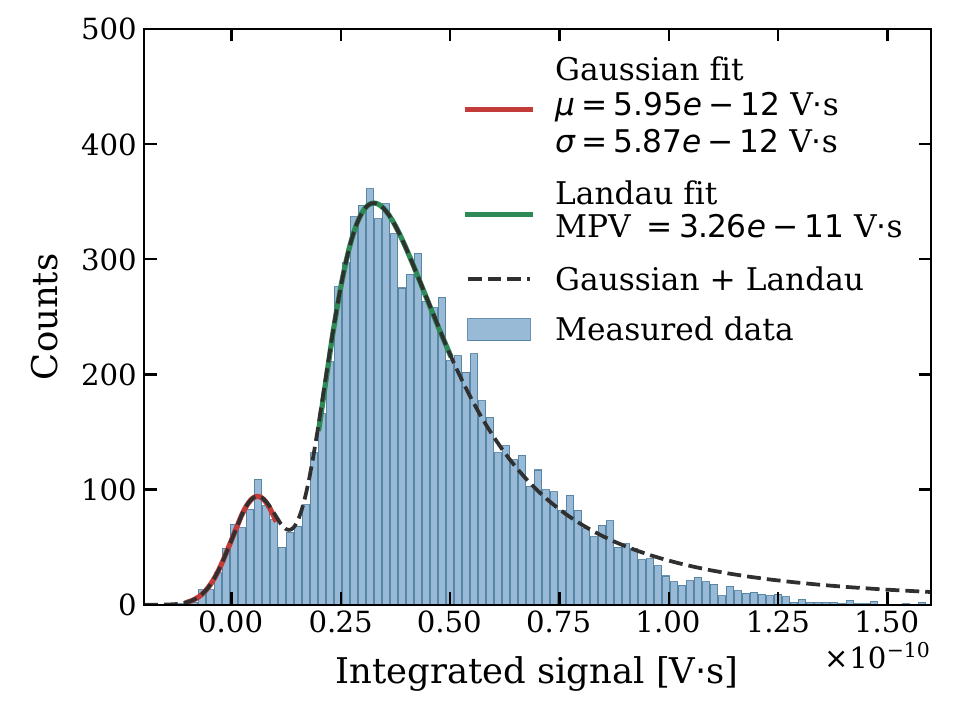}
    }

    \par\smallskip

    \subfloat[\label{fig:signal_event}]{%
        \includegraphics[width=0.43\linewidth]{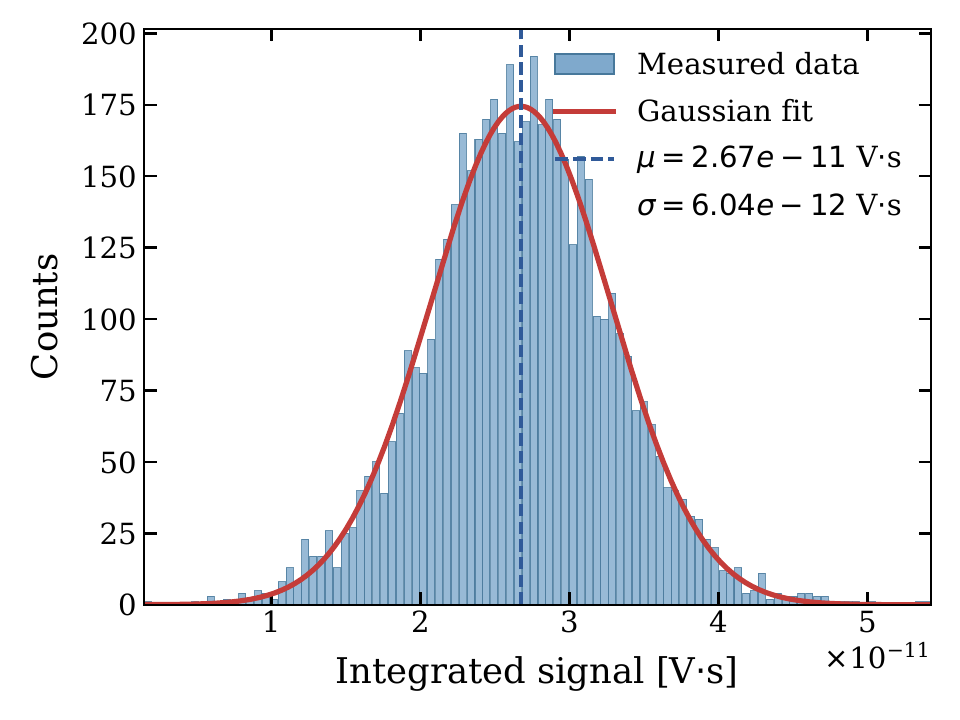}
    }
    \hfill
    \subfloat[\label{fig:average_pulse}]{%
        \includegraphics[width=0.43\linewidth]{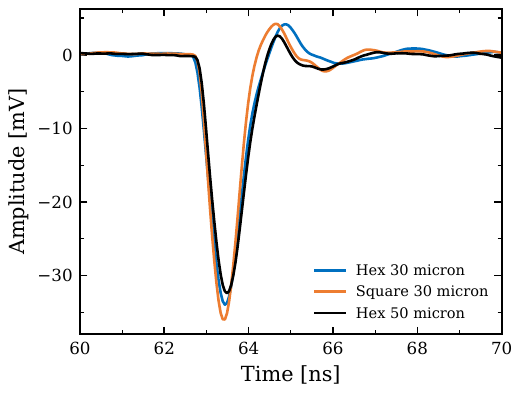}
    }

    \caption{$^{90}$Sr and laser calibration measurements.
    (a) Representative $^{90}$Sr waveforms and
    (b) the corresponding integrated-signal distribution
    for the Square30 sensor.
    (c) Integrated-signal distribution obtained from
    laser measurements.
    (d) Average laser-induced waveforms for the three
    sensor geometries at the scan position (-10\si{\micro\meter},0,0)
    indicated in Figure~\ref{fig:spatial_scan}.}

    \label{fig:calibration_timing}
\end{figure}

The laser intensity was calibrated using
$^{90}$Sr measurements to establish a MIP-like
signal reference.
The measurements were performed on the Square30
sensor at an applied voltage of \SI{50}{\volt},
corresponding to an effective sensor bias of
approximately \SI{40}{\volt} after correcting
for the voltage drop across the bias resistors.

Approximately $10^4$ waveforms were recorded
using the oscilloscope, which was triggered
directly by the detector signal at a threshold
of \SI{-20}{\milli\volt}.
For each waveform, the baseline mean and standard
deviation were calculated from the first 100 samples.
Waveforms with a baseline standard deviation of
\SI{5}{\milli\volt} or greater were rejected.
Since the detector signals were negative,
waveforms with positive voltage excursions exceeding
\SI{30}{\milli\volt} above the baseline were also
rejected to suppress abnormal waveforms affected by
oscillatory interference from an unidentified
external noise source.

\par

For each accepted waveform, the baseline was
subtracted and the signal polarity was reversed.
The signal was then integrated using the trapezoidal
rule over a time window of
$\pm\SI{2}{\nano\second}$ around the minimum of
the original waveform.
The resulting voltage integral, expressed in
$\mathrm{V}\cdot\mathrm{s}$, was used as a
relative measure of signal magnitude, since
the conversion factor between the integrated
output voltage and the collected charge had not
been independently calibrated.
The integrated-signal distribution was fitted
with a Gaussian component describing the
noise-dominated population and a Landau
component describing the $\beta$-particle
signals~\cite{b10}.
The most probable value (MPV) of the Landau
component, $3.26\times10^{-11}~\mathrm{V}\cdot\mathrm{s}$,
was used as the reference for laser-intensity calibration.
Representative results for the Square30 sensor
are shown in Figure~\ref{fig:calibration_timing}(a,b).


\par

To investigate the potential influence of the
oscilloscope trigger threshold on the extracted MPV,
an additional measurement was performed with a
trigger threshold of \SI{-18}{\milli\volt},
yielding an MPV of approximately
$2.848\times10^{-11}~\mathrm{V}\cdot\mathrm{s}$,
which differed by approximately 13\% from that
obtained with the nominal trigger threshold of
\SI{-20}{\milli\volt}.
However, only approximately 10\% of the waveforms
recorded at the \SI{-18}{\milli\volt} trigger
threshold were attributed to $\beta$-particle signals.
The sensitivity to the fitting range was evaluated
by varying its upper boundary from
$5\times10^{-11}$ to
$1\times10^{-10}~\mathrm{V}\cdot\mathrm{s}$,
while keeping the lower boundary fixed at
$2\times10^{-11}~\mathrm{V}\cdot\mathrm{s}$.
The extracted MPV varied by less than 1\% for the
\SI{-20}{\milli\volt} measurement, compared with
approximately 10\% for the \SI{-18}{\milli\volt}
measurement.
Consequently, the observed 13\% difference cannot
be unambiguously attributed to the trigger threshold
alone, given the limited signal statistics and the
greater sensitivity to the fitting range observed
for the \SI{-18}{\milli\volt} measurement.

\par
The laser-induced signals were processed using
the same baseline subtraction, polarity reversal,
and trapezoidal integration procedure as in the
$^{90}$Sr analysis, with the same integration window.
The resulting integrated-signal distribution was
fitted with a Gaussian function, as shown in
Figure~\ref{fig:calibration_timing}(c), yielding a mean
of $2.67\times10^{-11}~\mathrm{V}\cdot\mathrm{s}$.
The laser intensity was adjusted so that the
fitted mean was approximately 18\% below the
$^{90}$Sr Landau MPV obtained with the nominal
trigger threshold of \SI{-20}{\milli\volt},
to avoid overestimating the MIP-like signal magnitude.
\par
Following calibration, approximately 2000 recorded
laser-induced waveforms were averaged for each sensor.
The resulting pulse shapes are compared in
Figure~\ref{fig:calibration_timing}(d).
\par

Although the integrated laser signals are comparable
to the MIP-like reference, differences in pulse
shape, particularly in the leading-edge slope,
can influence the electronic timing jitter and,
consequently, the measured timing resolution.
Furthermore, laser injection does not reproduce
the spatial distribution of deposited charge or
the stochastic energy-deposition fluctuations
associated with charged particles traversing
the sensor.
Therefore, the laser measurements provide a
controlled comparison of the timing response
of the three sensor geometries rather than
a direct measurement of the timing resolution
for MIPs.

\section{Timing analysis}
\label{sec:timing_analysis}

\begin{figure}[htbp]
    \centering

    \subfloat[\label{fig:timing_cal}]{%
        \includegraphics[width=0.43\linewidth]
        {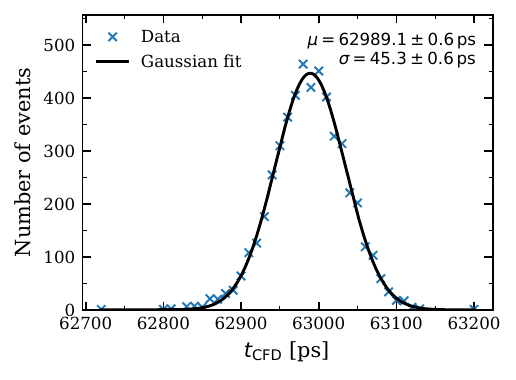}
    }
    \hfill
    \subfloat[\label{fig:threshold}]{%
        \includegraphics[width=0.43\linewidth]
        {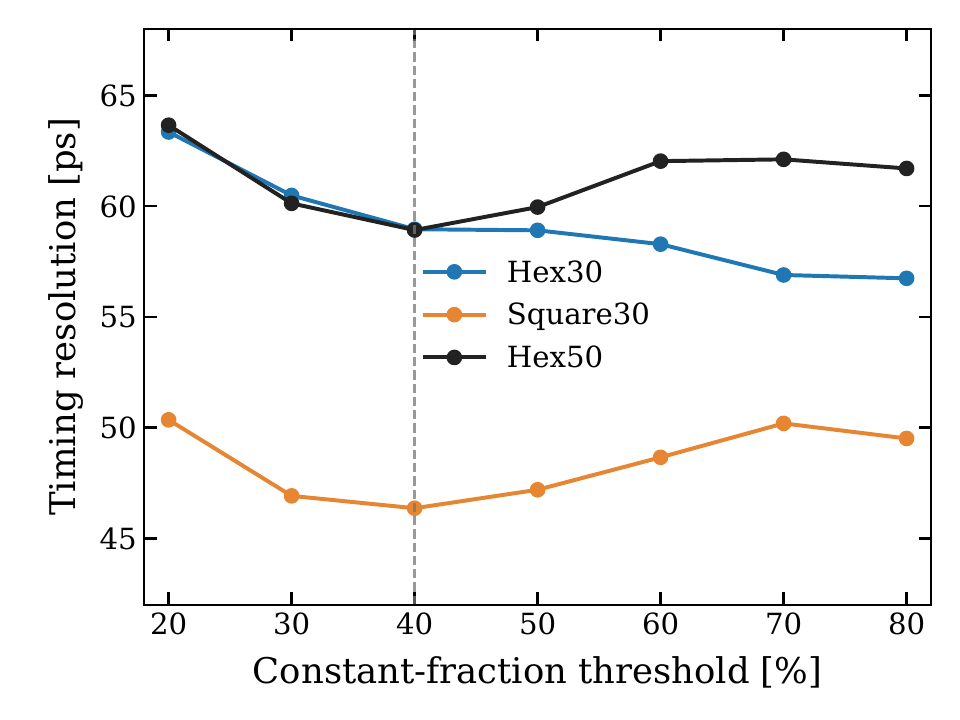}
    }

    \par\smallskip

    \subfloat[\label{fig:sigma_bias}]{%
        \includegraphics[width=0.43\linewidth]
        {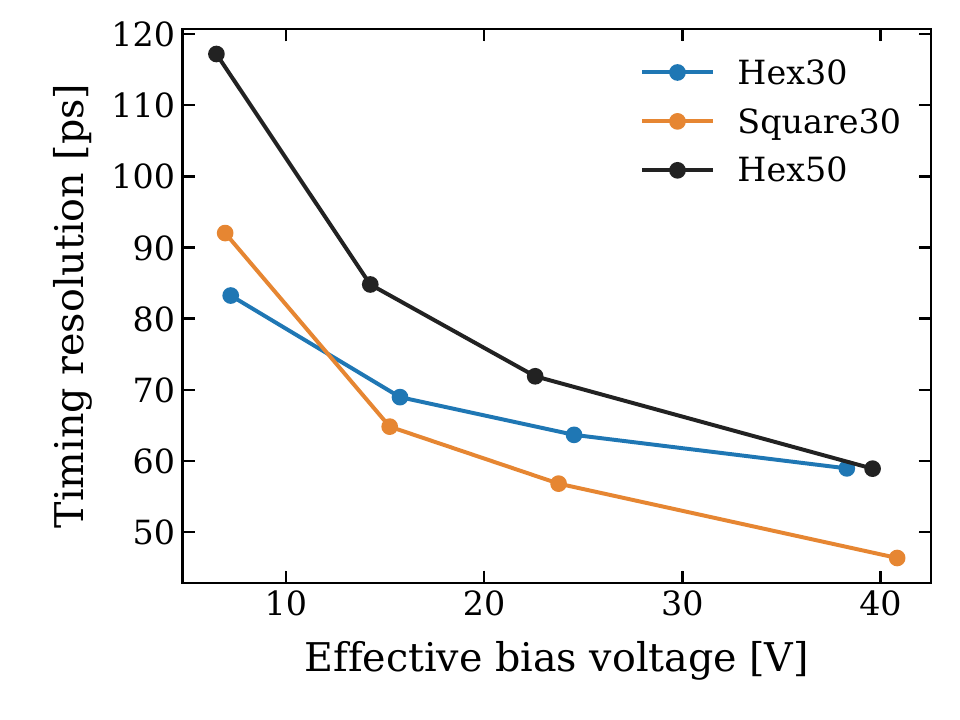}
    }
    \hfill
    \subfloat[\label{fig:mu_bias}]{%
        \includegraphics[width=0.43\linewidth]
        {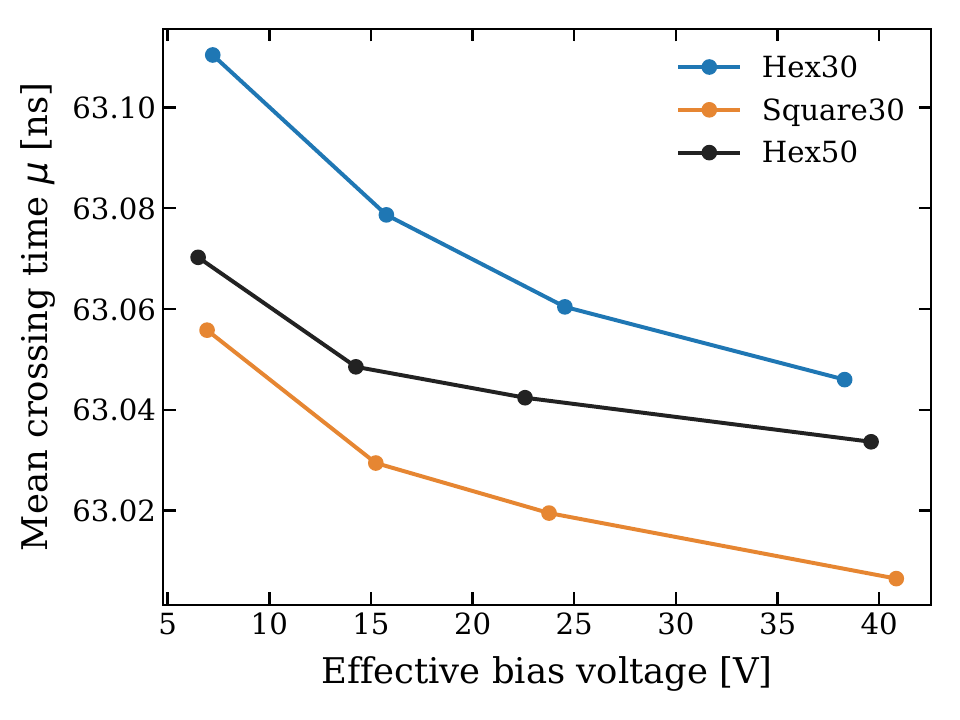}
    }

    \caption{Laser-based timing measurements at the
    scan position (-10\si{\micro\meter},0,0)
    indicated in Figure~\ref{fig:spatial_scan}.
    (a) Gaussian fit to the threshold-crossing-time
    distribution at a threshold fraction of 40\%.
    (b) Timing resolution as a function of threshold
    fraction for the three sensor geometries.
    (c) Timing resolution $\sigma$ and
    (d) mean threshold-crossing time $\mu$
    as functions of effective bias voltage,
    both extracted using a threshold fraction of 40\%.}

    \label{fig:timing_analysis}
\end{figure}

In this section, representative timing results
obtained at the fixed laser injection position
(-10\si{\micro\meter},0,0) are presented.
For each waveform, the threshold-crossing time
was extracted from the leading edge at a specified
fraction of its individual pulse amplitude.
The resulting threshold-crossing-time distribution
was fitted with a Gaussian function, as shown in
Figure~\ref{fig:timing_cal}.
The fitted standard deviation $\sigma$ was taken
as the measured local timing resolution
$\sigma_{\mathrm{local}}$ at the given injection
position, including the contributions described
in Equation~\ref{eq:timing_resolution}.

\par

The dependence of the timing resolution on the
threshold fraction is shown in
Figure~\ref{fig:threshold}.
For the Square30 and Hex50 sensors, the timing
resolution initially improves with increasing
threshold fraction, reaches a minimum near 40\%,
and subsequently deteriorates.
In contrast, the timing resolution of Hex30 generally
improves with increasing threshold fraction over
the investigated range.
This distinct behavior may be related to differences
in sensor capacitance, as Hex30 has a $5\times10$
cell matrix, whereas Square30 and Hex50 have
$5\times5$ cell matrices.
Sensor capacitance can affect the readout response,
leading-edge slope, and electronic noise, thereby
influencing the dependence of the timing resolution
on the threshold fraction.

\par

Although the optimum threshold fraction differs
among the three sensors, a common threshold
fraction of 40\% was adopted for all subsequent
timing analyses to ensure a consistent comparison.
The relative statistical uncertainties on the
extracted timing resolutions were typically
1--2\%, while the absolute statistical uncertainties
on the mean threshold-crossing times $\mu$ were
typically below \SI{1}{\pico\second}.

\par

In addition to the local timing resolution $\sigma$,
the Gaussian fit provides the mean threshold-crossing
time $\mu$ relative to the timing reference.
While $\sigma$ characterizes event-to-event timing
fluctuations at a fixed injection position,
$\mu$ characterizes the average timing response,
which depends on carrier transport, signal formation,
and the readout response.

\par

Both quantities exhibit a dependence on the
effective bias voltage, as shown in
Figures~\ref{fig:sigma_bias} and~\ref{fig:mu_bias}.
The observed bias dependence is consistent with
faster charge collection and a steeper signal
leading edge at higher bias voltages, contributing
to improved timing performance.

\par

While $\sigma$ characterizes the local timing
resolution, spatial variations in $\mu$ introduce
an additional timing uncertainty when the particle
impact position is unknown.
Therefore, both the spatial distributions of
$\sigma$ and $\mu$ are relevant to the overall
timing performance of 3D sensors and are
investigated in the following section.

\section{Spatial uniformity}
\label{sec:spatial_uniformity}

\begin{figure}[htbp]
    \centering

    \subfloat[\label{fig:hex30_sigma_map}]{%
        \includegraphics[width=0.44\linewidth]
        {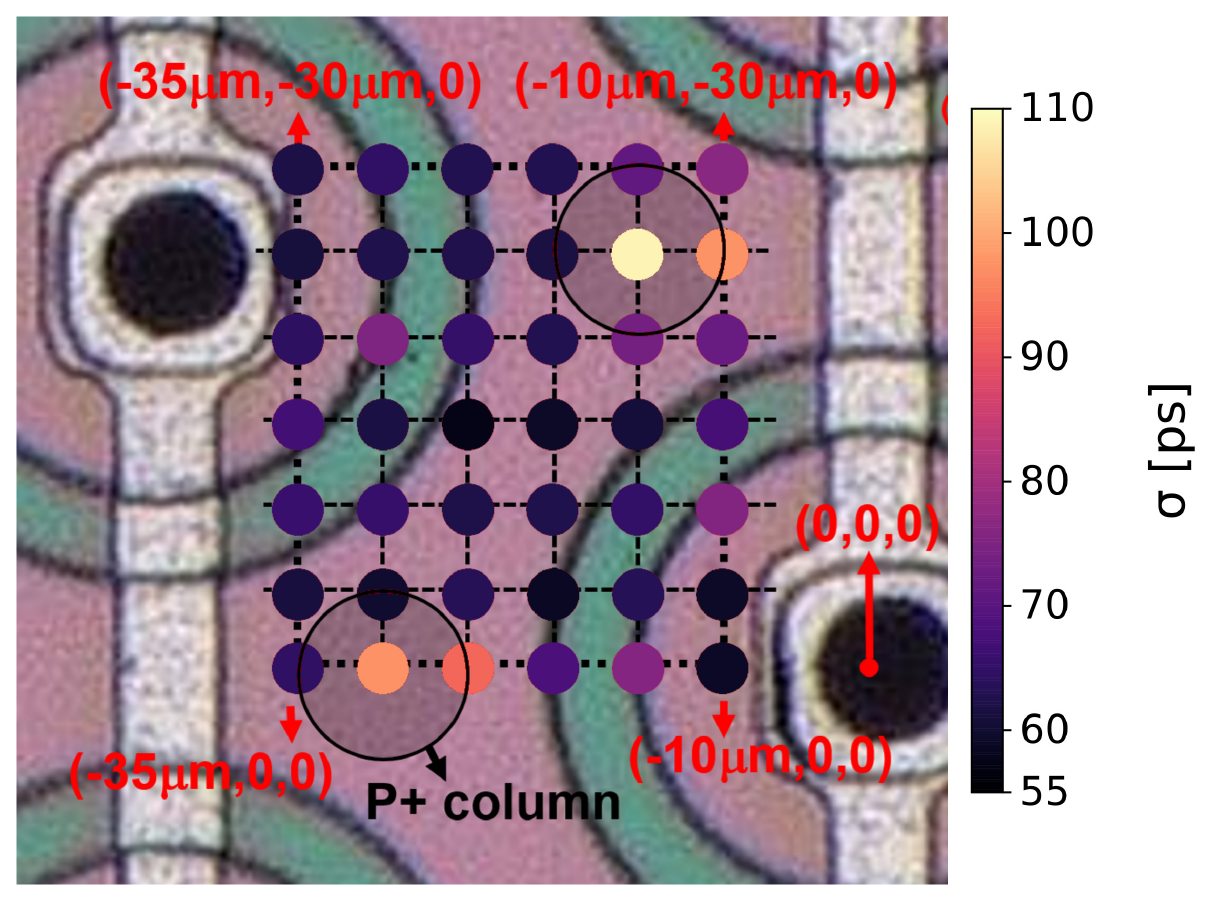}
    }
    \hfill
    \subfloat[\label{fig:hex30_mu_map}]{%
        \includegraphics[width=0.44\linewidth]
        {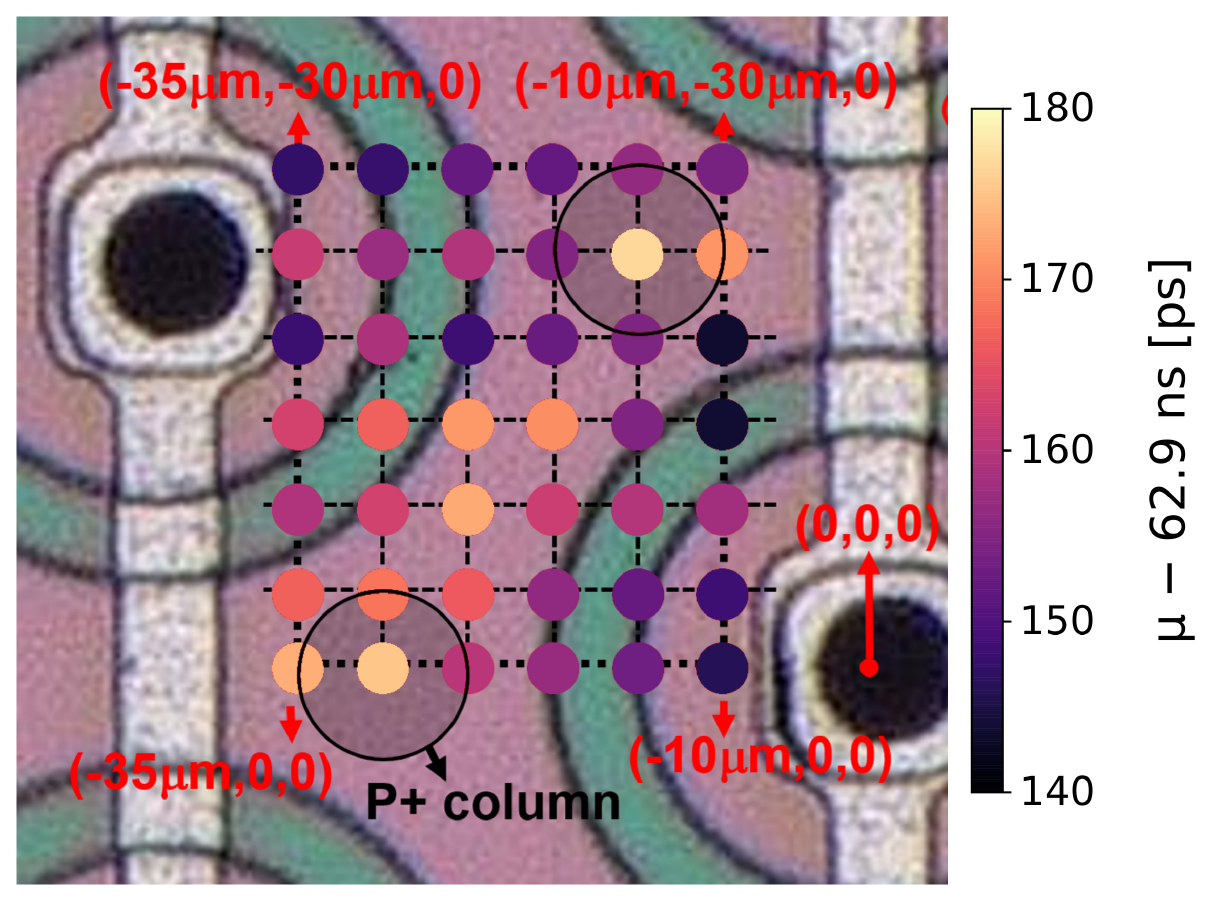}
    }

    \par\medskip

    \subfloat[\label{fig:square30_sigma_map}]{%
        \includegraphics[width=0.44\linewidth]
        {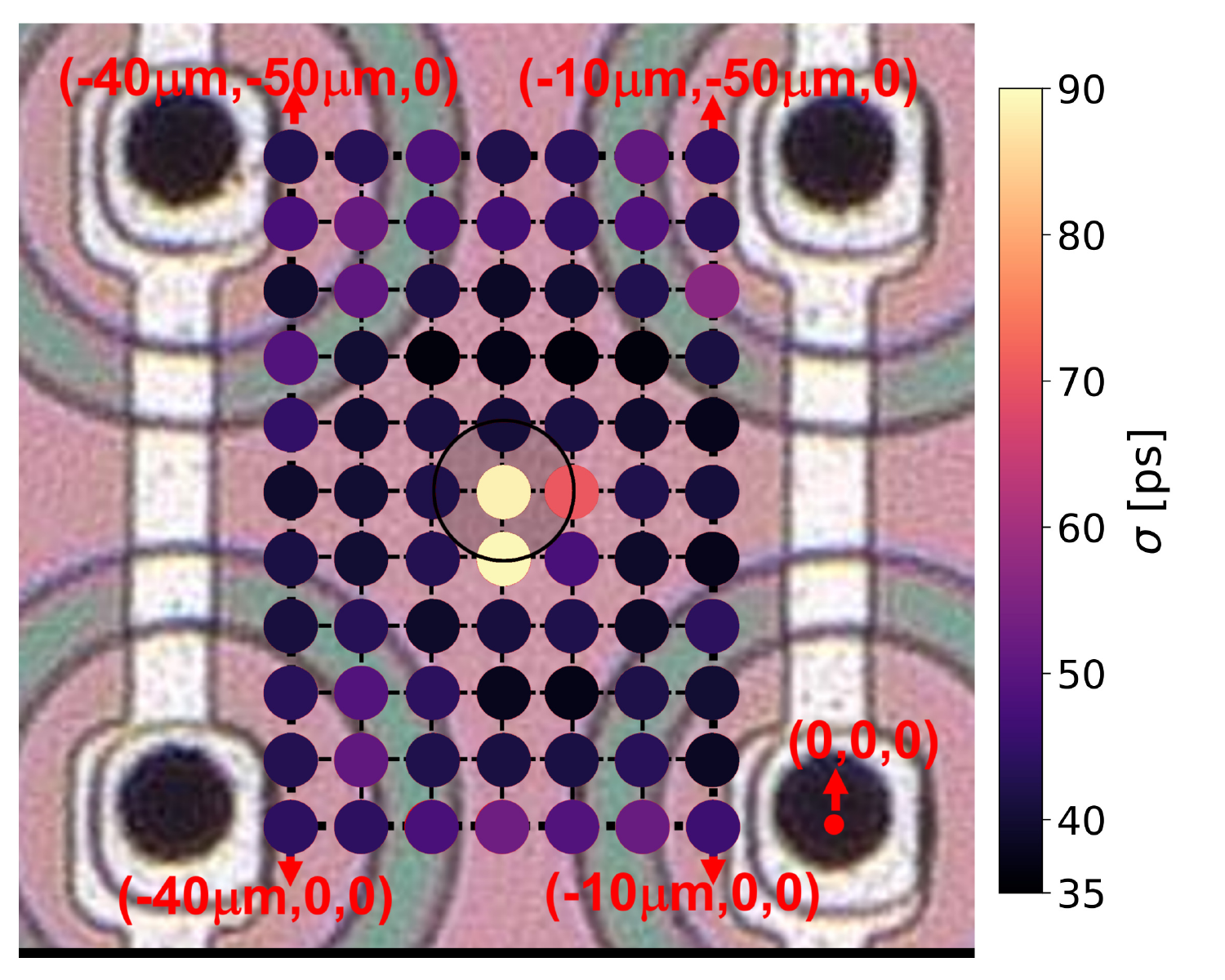}
    }
    \hfill
    \subfloat[\label{fig:square30_mu_map}]{%
        \includegraphics[width=0.44\linewidth]
        {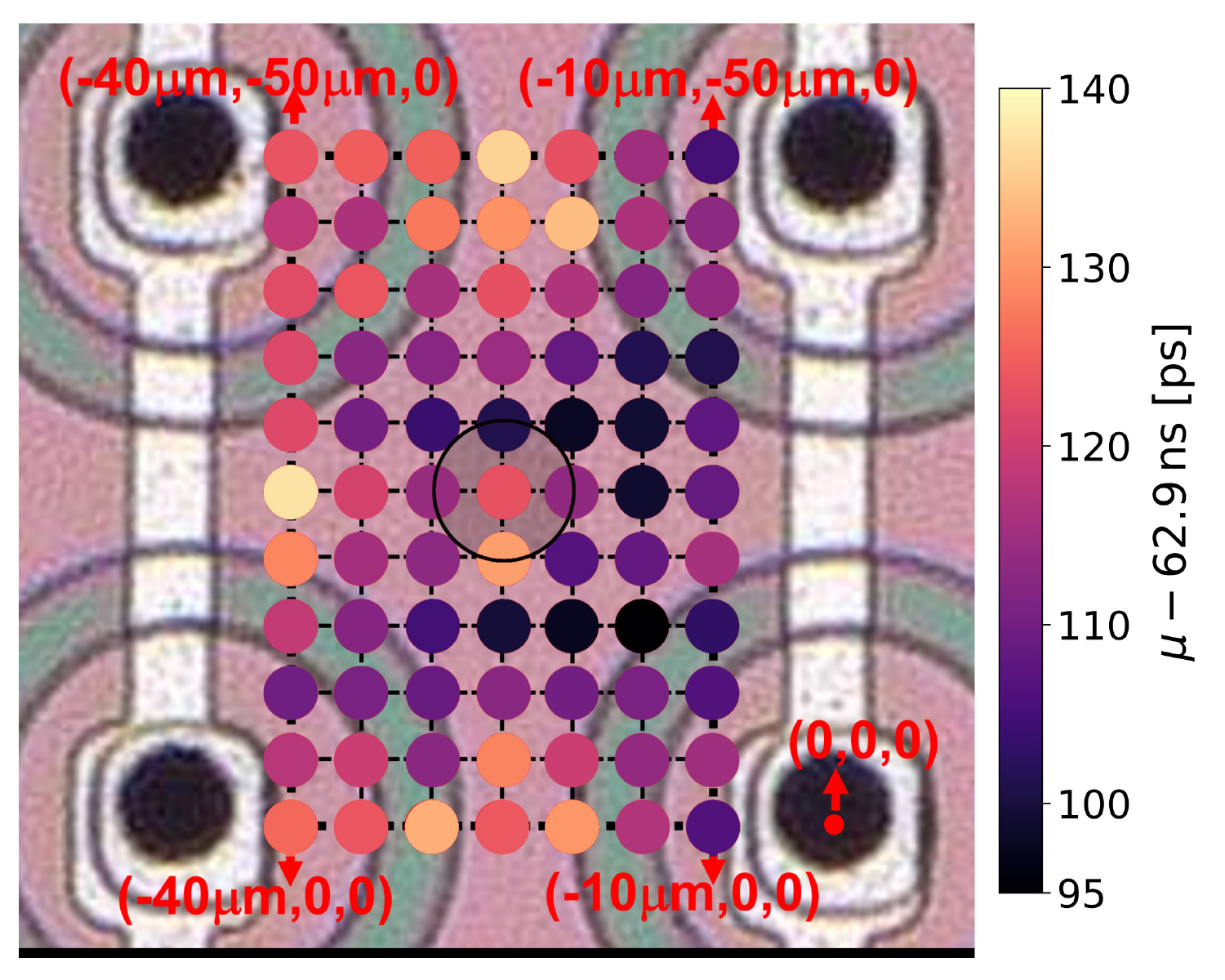}
    }

    \par\medskip

    \subfloat[\label{fig:hex50_sigma_map}]{%
        \includegraphics[width=0.44\linewidth]
        {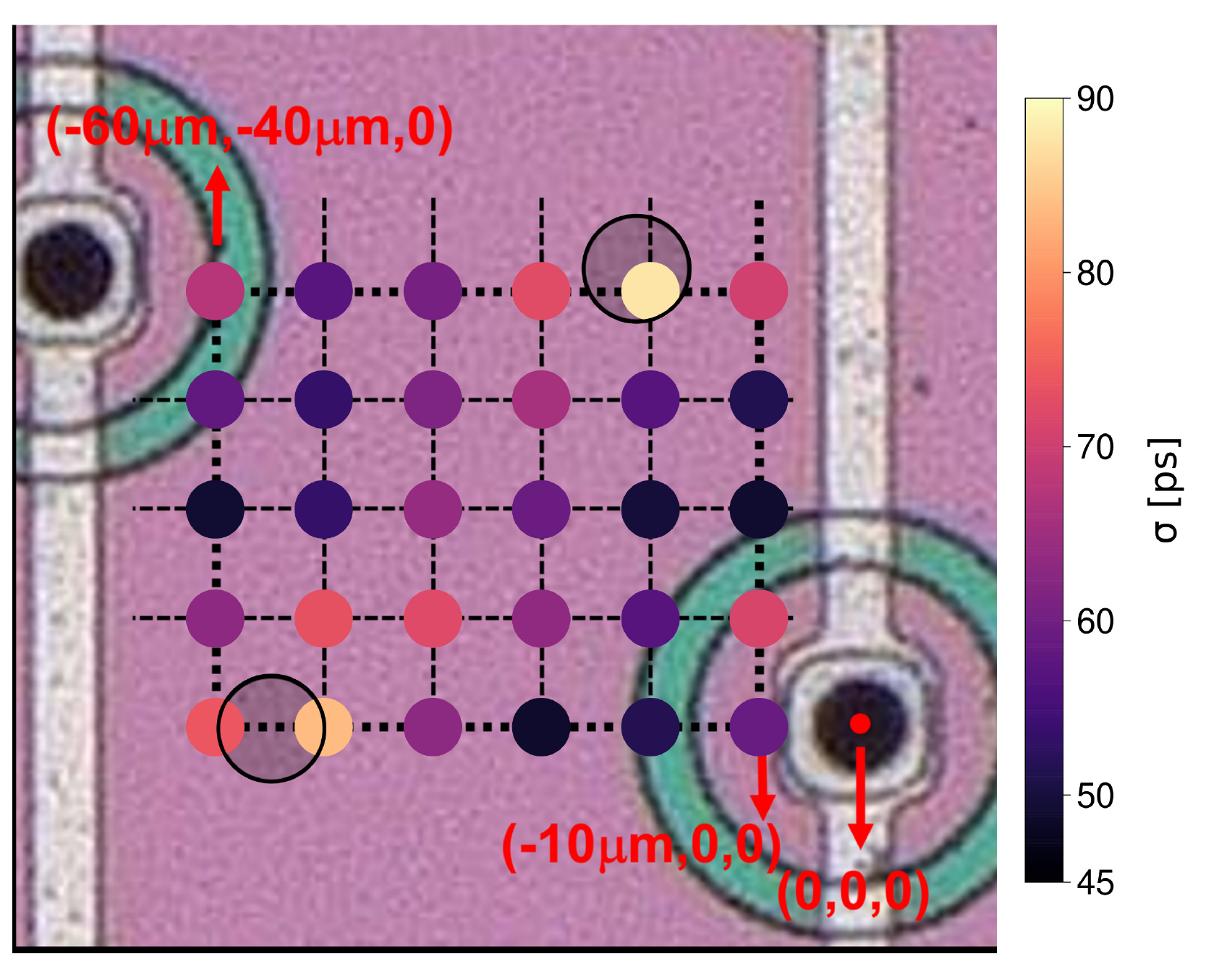}
    }
    \hfill
    \subfloat[\label{fig:hex50_mu_map}]{%
        \includegraphics[width=0.44\linewidth]
        {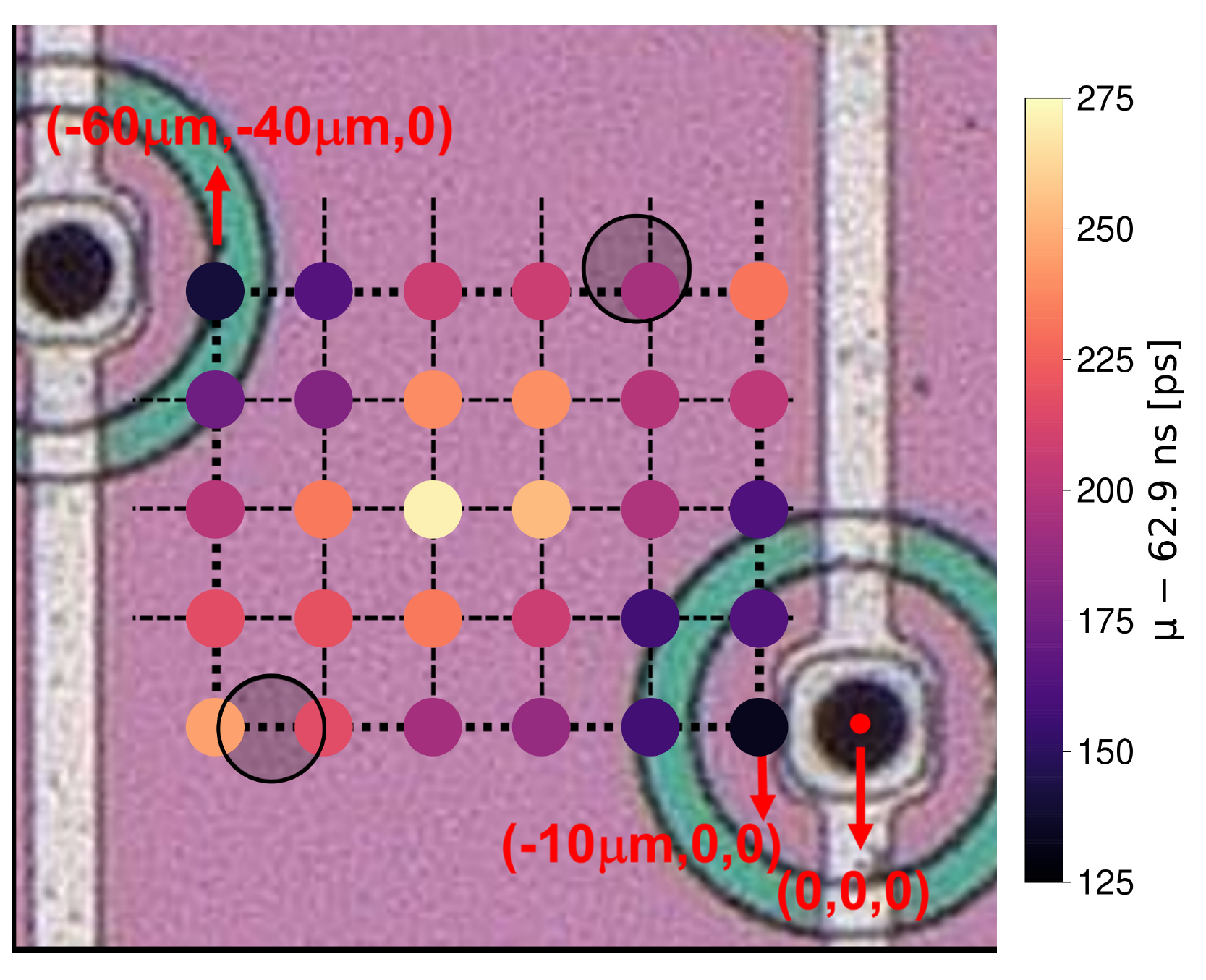}
    }

    \caption{Spatial distributions of the local timing
    resolution ($\sigma$, left) and the mean
    threshold-crossing time ($\mu$, right) for
    (a,b) Hex30, (c,d) Square30, and (e,f) Hex50,
    measured at an applied voltage of
    $V_{\mathrm{applied}}=\SI{50}{\volt}$ using a
    constant-fraction threshold of 40\%.
    A common time offset of \SI{62.9}{\nano\second}
    was subtracted from $\mu$ for visualization.
    The colored circles indicate the laser-scan
    positions, with the corresponding values expressed
    in picoseconds.
    The scan steps were \SI{5}{\micro\meter} for
    Hex30 and Square30 and \SI{10}{\micro\meter}
    for Hex50.
    The scan coordinates are defined relative to the
    reference $n^{+}$ column.
    The black circular outlines indicate the estimated
    positions of the backside $p^{+}$ columns,
    with a positional uncertainty of approximately
    1--2~\si{\micro\meter}.}

    \label{fig:spatial_scan}
\end{figure}

The spatial uniformity of the timing response was
investigated in terms of the local timing resolution
$\sigma$ and the mean threshold-crossing time $\mu$,
both extracted using a constant-fraction threshold
of 40\%.

Figure~\ref{fig:spatial_scan} shows the spatial
distributions of $\sigma$ and $\mu$ for the three
sensor geometries at an applied voltage of
\SI{50}{\volt}.
For Hex30 and Square30, the local timing resolution
is relatively uniform except at some positions
close to the estimated $p^{+}$ columns.
Hex50 exhibits more pronounced spatial variations,
with $\sigma$ increasing as the injection position
moves from the $n^{+}$ towards the $p^{+}$ columns.
This variation becomes less pronounced with
increasing bias voltage.
Position-dependent shifts in $\mu$ are also
observed for all three geometries.


To compare the three geometries using a common
spatial sampling step of \SI{10}{\micro\meter},
every second measurement point along both spatial
directions was retained from the original
\SI{5}{\micro\meter} scans of Hex30 and Square30.
The original sampling grid of Hex50 was preserved.
The differences between the spatial statistics
obtained using the two sampling steps were less
than 3\% for Hex30 and Square30.

The spatial averages and spatial standard deviations
(spatial STDs) of $\sigma$ and $\mu$ were calculated as

\begin{equation}
    \overline{q} =
    \frac{1}{N}\sum_{i=1}^{N}q_i,
    \qquad q \in \{\sigma,\mu\},
    \label{eq:spatial_mean}
\end{equation}

and

\begin{equation}
    s_q =
    \sqrt{
        \frac{1}{N}
        \sum_{i=1}^{N}
        (q_i-\overline{q})^2
    },
    \qquad q \in \{\sigma,\mu\},
    \label{eq:spatial_variation}
\end{equation}

where $q_i$ is the corresponding Gaussian fit
parameter at scan position $i$, and $N$ is the
number of selected scan positions.
The quantities $s_{\sigma}$ and $s_{\mu}$
are the spatial STDs of the local timing resolution
and the mean threshold-crossing time, respectively.


\begin{figure}[htbp]
    \centering

    \subfloat[\label{fig:mean_timing}]{%
        \includegraphics[width=0.44\linewidth]
        {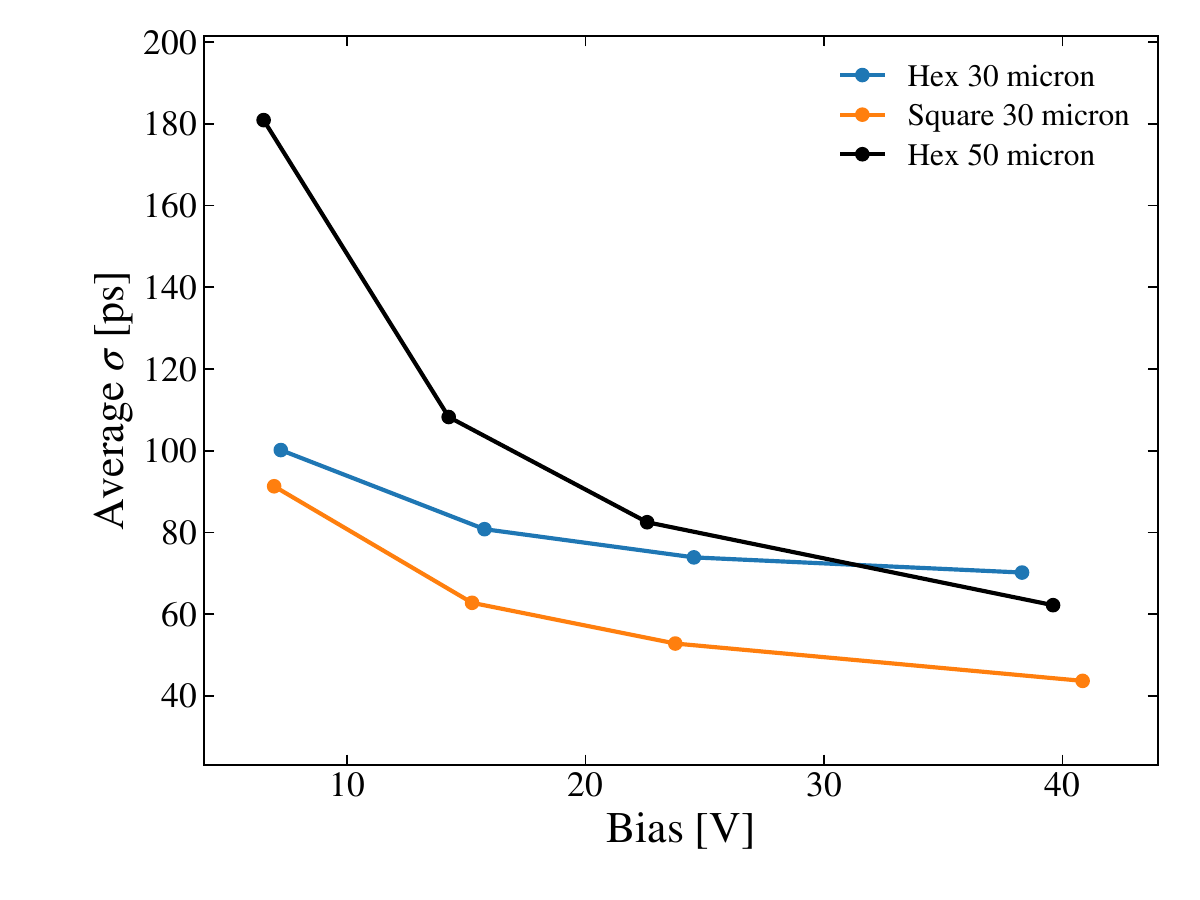}
    }
    \hfill
    \subfloat[\label{fig:std_timing}]{%
        \includegraphics[width=0.44\linewidth]
        {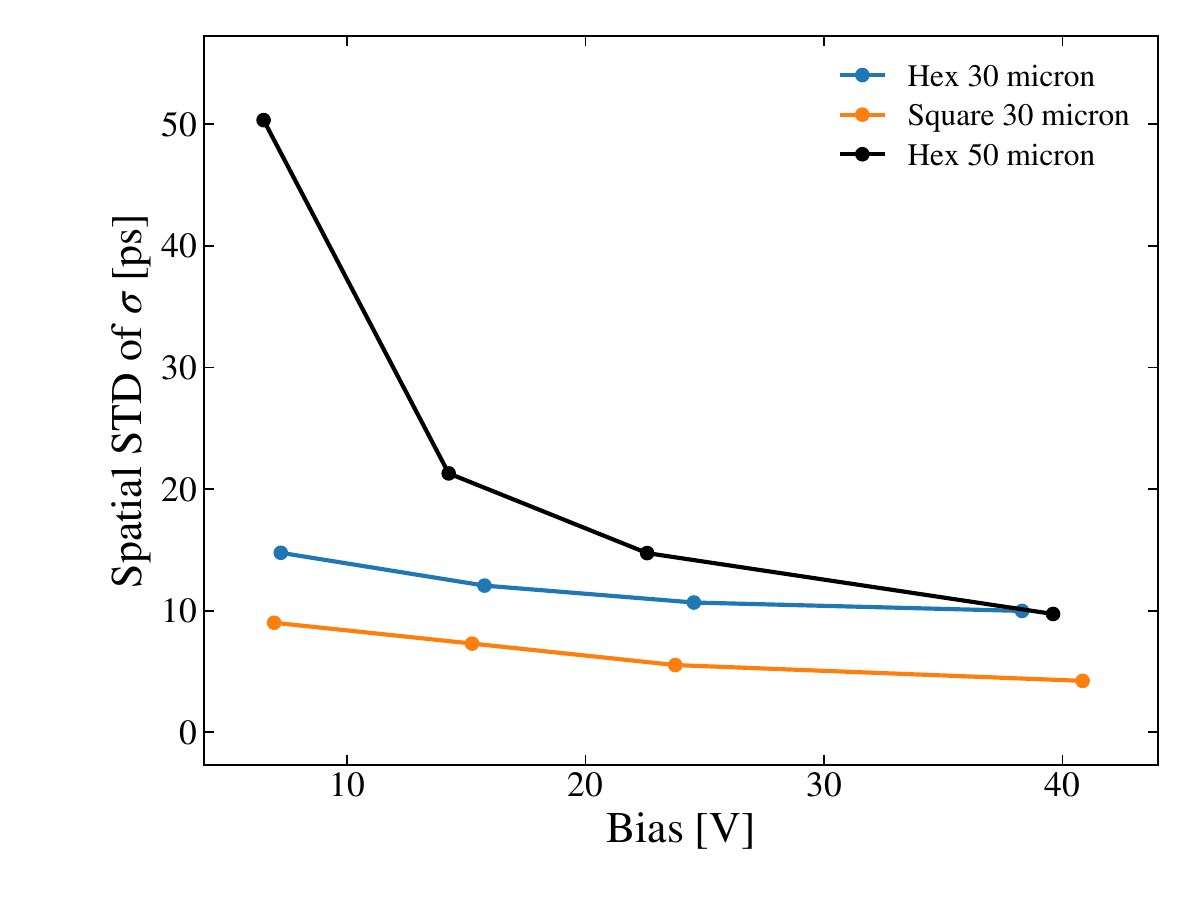}
    }

    \par\medskip

    \subfloat[\label{fig:mean_mu}]{%
        \includegraphics[width=0.44\linewidth]
        {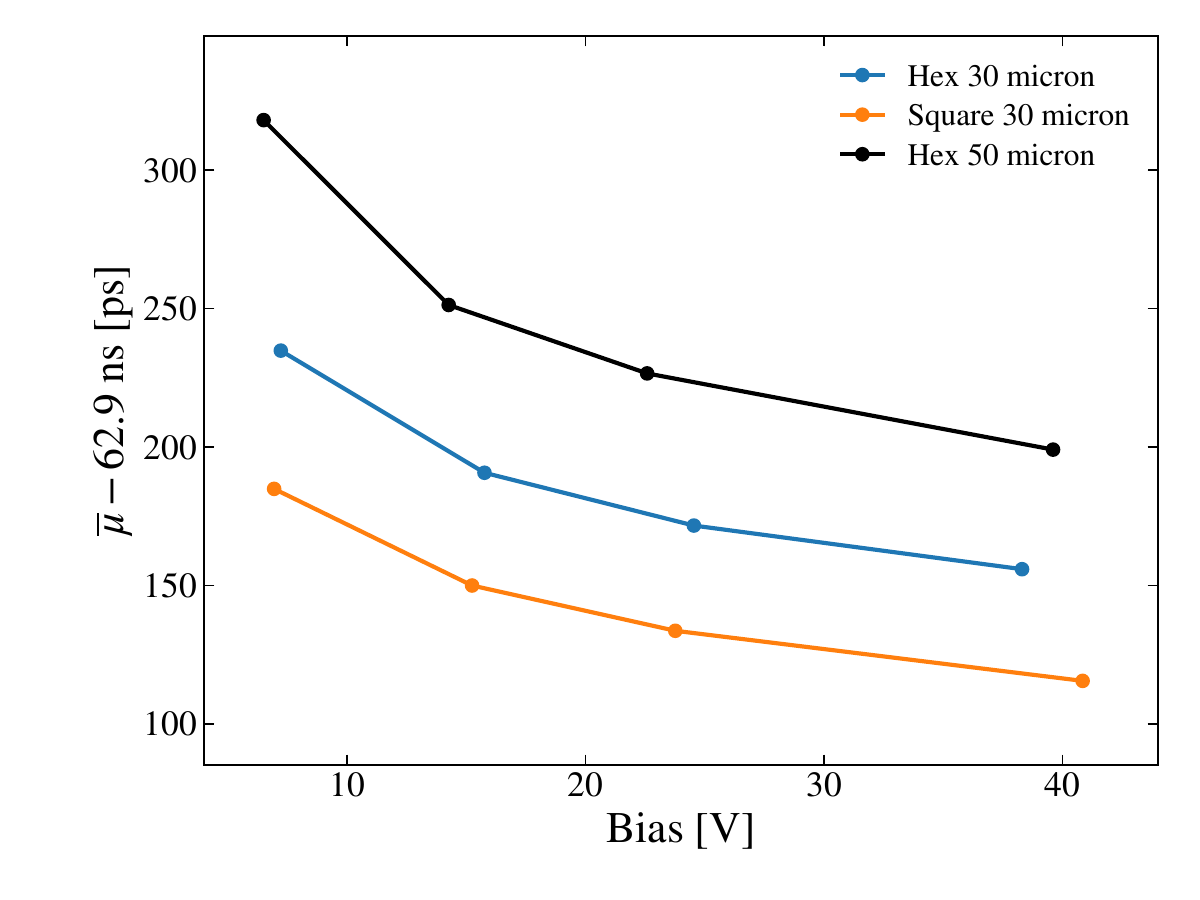}
    }
    \hfill
    \subfloat[\label{fig:std_mu}]{%
        \includegraphics[width=0.44\linewidth]
        {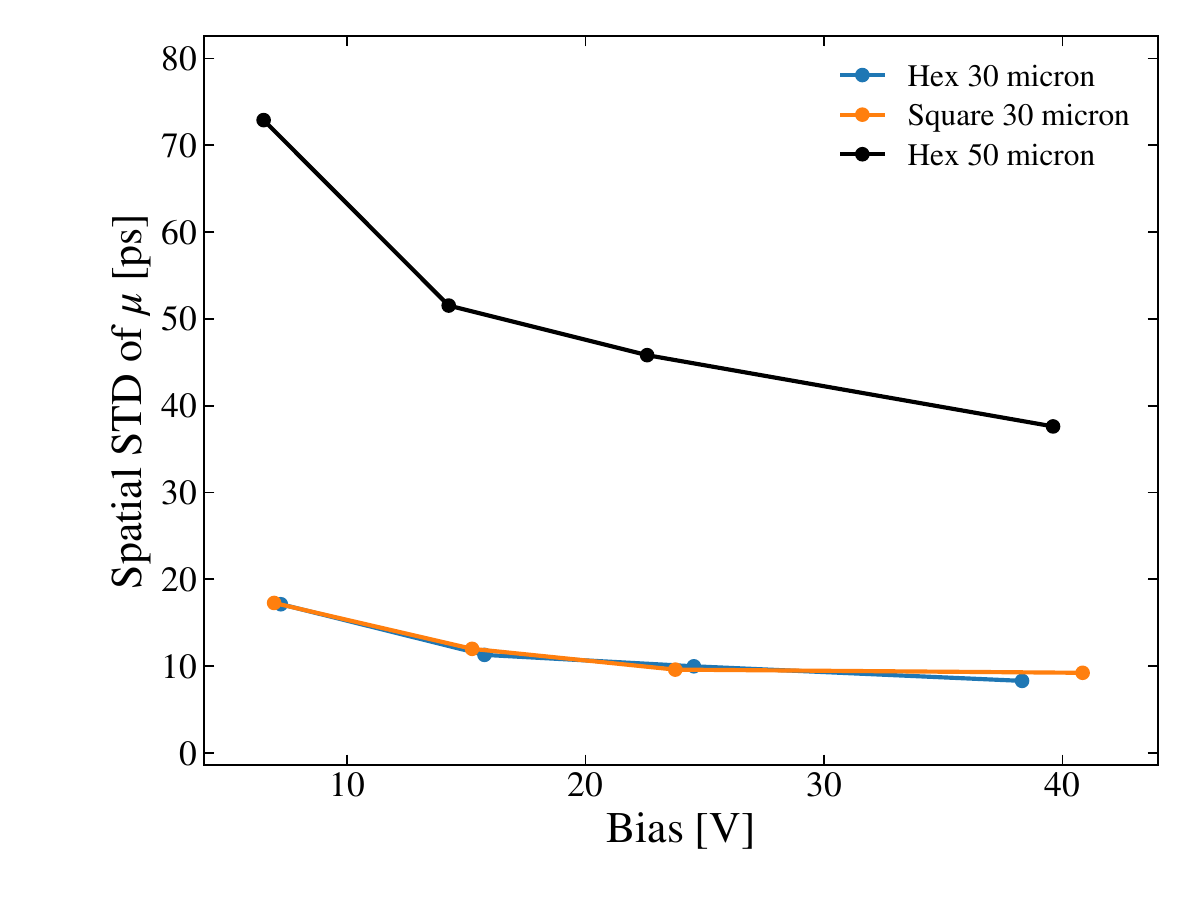}
    }

    \caption{Dependence on the effective sensor bias of
    (a) the average local timing resolution
    $\overline{\sigma}$,
    (b) its spatial standard deviation $s_{\sigma}$,
    (c) the spatial average of the mean threshold-crossing
    time $\overline{\mu}$ after subtraction of a common
    time offset of \SI{62.9}{\nano\second}, and
    (d) its spatial standard deviation $s_{\mu}$.
    All quantities were evaluated using a common
    spatial sampling step of \SI{10}{\micro\meter}.
    The spatial STDs characterize
    position-dependent variations rather than
    statistical uncertainties on the fitted parameters.}

    \label{fig:timing_summary}
\end{figure}

Figure~\ref{fig:timing_summary}(a) shows that
$\overline{\sigma}$ decreases with increasing
effective bias for all three geometries.

At low bias, Hex30 achieves an average local
timing resolution of approximately
\SI{100}{\pico\second} at \SI{7.2}{\volt},
compared with \SI{181}{\pico\second} for Hex50
at \SI{6.5}{\volt}.
At the highest measured bias, the corresponding
values are approximately \SI{70}{\pico\second}
and \SI{62}{\pico\second}, respectively.
The stronger bias dependence of Hex50 is
consistent with its larger inter-electrode
spacing and lower characteristic electric
field at low bias.
However, as shown in Figure~\ref{fig:threshold},
the common threshold fraction of 40\% is not
optimal for Hex30, which should be taken into
account when comparing the spatial averages.

For the two geometries with the same nominal
inter-electrode spacing, Square30 exhibits
better average local timing resolution than
Hex30 within the scanned regions.
The value for Square30 improves from approximately
\SI{91}{\pico\second} at \SI{6.9}{\volt}
to \SI{44}{\pico\second} at \SI{40.8}{\volt}.

As shown in Figure~\ref{fig:timing_summary}(b),
$s_{\sigma}$ decreases with increasing bias
for all three geometries.
For Hex50, $s_{\sigma}$ decreases from
approximately \SI{50}{\pico\second}
to \SI{10}{\pico\second}.
The corresponding values decrease from
\SI{15}{\pico\second} to \SI{10}{\pico\second}
for Hex30 and from \SI{9}{\pico\second}
to \SI{4}{\pico\second} for Square30.
Square30 exhibits the smallest spatial variation
of $\sigma$ over the investigated bias range.

Figure~\ref{fig:timing_summary}(c) shows that
$\overline{\mu}$ decreases with increasing bias
for all three geometries.
Differences in $\overline{\mu}$ among the sensors
may also arise from differences in the readout boards
and wire-bond connections.

As shown in Figure~\ref{fig:timing_summary}(d),
$s_{\mu}$ also decreases with increasing bias.
For Hex30 and Square30, $s_{\mu}$ decreases
from approximately \SI{17}{\pico\second}
to 8--9~\si{\pico\second}, while for
Hex50 it decreases from approximately
\SI{73}{\pico\second} to \SI{38}{\pico\second}.
The two \SI{30}{\micro\meter} geometries
exhibit smaller spatial variations in $\mu$
than Hex50 over the investigated bias range.


For unknown impact positions, the spatial
variation of $\mu$ introduces an additional
timing uncertainty.
Assuming a uniform probability distribution
of particle impact positions within the scanned
region, approximated by equal weights for the selected
scan positions, and no position-dependent timing correction,
the combined timing resolution can be estimated
using the law of total variance as

\begin{equation}
    \sigma_{\mathrm{combined}}^2
    =
    \frac{1}{N}\sum_{i=1}^{N}\sigma_i^2
    + s_{\mu}^2
    =
    \overline{\sigma}^{\,2}
    + s_{\sigma}^2
    + s_{\mu}^2,
    \label{eq:combined_timing}
\end{equation}

where $\sigma_i$ is the measured local timing
resolution at scan position $i$,
$\overline{\sigma}$ is its spatial average,
$s_{\sigma}$ is the spatial STD of the local timing
resolution,
and $s_{\mu}$ is the spatial STD of the mean
threshold-crossing time.

The timing jitter associated with the
electronics-only noise was estimated using
Equation~\ref{eq:electronic_jitter}, with an RMS output voltage
noise of approximately \SI{0.5}{\milli\volt}
measured without the sensor connected.
The signal slope at the 40\% threshold was
obtained from the average pulse measured at
(-10\si{\micro\meter},0,0) for each sensor.

Based on Equation~\ref{eq:timing_resolution},
the corrected local timing resolution at
scan position $i$ is estimated as

\begin{equation}
    \sigma_{\mathrm{corr},i}^2
    \simeq
    \sigma_i^2
    - \sigma_{\mathrm{jitter,elec}}^2
    - \sigma_{\mathrm{ref}}^2,
    \label{eq:corrected_local_timing}
\end{equation}

where $\sigma_{\mathrm{ref}}$ is approximately
\SI{5}{\pico\second}.
Assuming that the estimated electronics-only jitter
is representative of the selected scan positions,
the corrected combined timing resolution is
given by

\begin{equation}
\begin{aligned}
    \sigma_{\mathrm{combined,corr}}^2
    &\simeq
    \frac{1}{N}\sum_{i=1}^{N}
    \sigma_{\mathrm{corr},i}^2
    + s_{\mu}^2 \\
    &\simeq
    \overline{\sigma}^{\,2}
    + s_{\sigma}^2
    + s_{\mu}^2
    - \sigma_{\mathrm{jitter,elec}}^2
    - \sigma_{\mathrm{ref}}^2 .
\end{aligned}
    \label{eq:corrected_combined_timing}
\end{equation}

The correction accounts for the estimated
electronics-only and reference jitter, but
not the additional noise observed with the
sensors connected. It therefore does not
represent the intrinsic sensor timing resolution.
\par
The contributions of random electronics and
reference jitter to the statistical uncertainty
of $\mu$ are negligible compared with its
observed spatial variation.

The measured timing performance and estimated
timing contributions at the highest investigated
bias are summarized in
Table~\ref{tab:timing_summary}.

\begin{table}[htbp]
    \centering

    \caption{Summary of the measured timing performance
    and estimated timing contributions at an applied
    voltage of \SI{50}{\volt} using a constant-fraction
    threshold of 40\%.
    Here, $s_{\sigma}$ and $s_{\mu}$ denote the
    spatial STDs of $\sigma$ and $\mu$, respectively.
    The electronics-only timing jitter was estimated
    using the average pulses measured at
    $(-10,0,0)$~\si{\micro\meter}.}

    \label{tab:timing_summary}

    \begin{tabular}{lccc}
        \hline
        Parameter & Hex30 & Square30 & Hex50 \\
        \hline

        $\left|\mathrm{d}V/\mathrm{d}t\right|_{40\%}$
        [mV/ns]
        & 63.5 & 80.6 & 66.7 \\

        $\sigma_{\mathrm{jitter,elec}}$ [ps]
        & 7.9 & 6.2 & 7.5 \\

        $\sigma_{\mathrm{ref}}$ [ps]
        & 5 & 5 & 5 \\

        \hline

        $\overline{\sigma}$ [ps]
        & 70 & 44 & 62 \\

        $s_{\sigma}$ [ps]
        & 10 & 4 & 10 \\

        $s_{\mu}$ [ps]
        & 8.3 & 9.2 & 38 \\

        \hline

        $\sigma_{\mathrm{combined}}$ [ps]
        & 71.2 & 45.1 & 73.4 \\

        $\sigma_{\mathrm{combined,corr}}$ [ps]
        & 70.6 & 44.4 & 72.8 \\

        \hline
    \end{tabular}
\end{table}

\section{Conclusion}
\label{sec:conclusion}

The timing performance and spatial uniformity
of three double-sided $p$-type 3D silicon
sensor layouts produced by CNM were investigated
using a \SI{1064}{\nano\meter} infrared laser,
with the laser signal calibrated against
a $^{90}$Sr reference.

The average local timing resolution improves
with increasing bias voltage for all three
sensor geometries.
At low bias, Hex30 achieves an average local
timing resolution of approximately
\SI{100}{\pico\second}, compared with
\SI{181}{\pico\second} for Hex50.
Square30 exhibits the best average local
timing resolution within the scanned regions,
improving from approximately
\SI{91}{\pico\second} at \SI{6.9}{\volt}
to \SI{44}{\pico\second} at \SI{40.8}{\volt}.

The spatial variations of both the local timing
resolution $\sigma$ and the mean threshold-crossing
time $\mu$ decrease with increasing bias voltage.
Square30 exhibits the smallest spatial variation
of $\sigma$, while Hex50 exhibits a substantially
larger spatial variation of $\mu$ than the two
\SI{30}{\micro\meter} geometries.

After accounting for the spatial variations
and correcting for the estimated electronics-only
and laser reference jitter, the corrected combined
timing resolutions at the highest investigated bias
are approximately \SI{71}{\pico\second},
\SI{44}{\pico\second}, and
\SI{73}{\pico\second} for Hex30, Square30,
and Hex50, respectively.
These values should be regarded as approximate
estimates, since the correction does not fully
account for the noise contributions with the
sensors connected or the spatial dependence
of the signal slope.
They therefore do not represent the intrinsic
timing resolutions of the sensors.
Overall, the results suggest that both
electrode spacing and arrangement are important
for the timing performance and spatial uniformity
of 3D silicon sensors.

In this study, the mean integrated laser signal
was approximately 18\% below the $^{90}$Sr
reference MPV.
Future investigations will include direct
timing measurements with charged particles
to establish the timing performance for MIPs,
measurements after irradiation, and detailed
electric-field and weighting-field simulations
to guide the optimization of electrode
geometry and spacing for radiation-hard
3D silicon timing detectors.

\acknowledgments

This work was supported in part by the Spanish Ministry of
Science, Innovation and Universities through the Nuclear
and Particle Physics programme under project
PID2024-158059NB-C32.
This work also made use of the Spanish ICTS Network
MICRONANOFABS, partially supported by MINECO.


\end{document}